\documentclass[obeyspaces,
  journal=pasa,
  manuscript=research-paper, 
  year=202X,
  volume=YY,
]{cup-journal}

\usepackage{amsmath}
\usepackage{amssymb,microtype,siunitx,booktabs}
\usepackage[skip=0.5ex]{subcaption}
\usepackage{url}
\usepackage{hyperref} 

\title{The Evolutionary Map of the Universe: A new radio atlas for the southern hemisphere sky}

\author{A. M. Hopkins}
\affiliation{School of Mathematical and Physical Sciences, 12 Wally's Walk, Macquarie University, NSW 2109, Australia}
\email[A. M. Hopkins]{andrew.hopkins@mq.edu.au}

\author{A. Kapinska}
\affiliation{National Radio Astronomy Observatory, PO Box 0, Socorro, NM 87801, USA}

\author{J. Marvil}
\affiliation{National Radio Astronomy Observatory, PO Box 0, Socorro, NM 87801, USA}

\author{T. Vernstrom}
\affiliation{Australia Telescope National Facility, CSIRO, Space and Astronomy, PO Box 1130, Bentley, WA 6151, Australia}
\alsoaffiliation{ICRAR, The University of Western Australia, 35 Stirling Hwy, 6009 Crawley, Australia}

\author{J. D. Collier}
\affiliation{Australian SKA Regional Centre, Curtin Institute of Radio Astronomy (CIRA), 1 Turner Avenue, Technology Park, Bentley, Western Australia, 6102}
\alsoaffiliation{Inter-University Institute for Data Intensive Astronomy (IDIA),
Department of Astronomy, University of Cape Town, South Africa}
\alsoaffiliation{Western Sydney University, Locked Bag 1797, Penrith South DC, NSW 2751, Australia}

\author{R. P. Norris}
\affiliation{Australia Telescope National Facility, CSIRO, Space and Astronomy, P.O. Box 76, Epping, NSW 1710, Australia}
\alsoaffiliation{Western Sydney University, Locked Bag 1797, Penrith South DC, NSW 2751, Australia}

\author{Y. A. Gordon}
\affiliation{Department of Physics, University of Wisconsin-Madison, 1150 University Avenue, Madison, WI 53706, USA}

\author{S.~W.~Duchesne}
\affiliation{Australia Telescope National Facility, CSIRO, Space and Astronomy, PO Box 1130, Bentley, WA 6151, Australia}

\author{L. Rudnick}
\affiliation{University of Minnesota, 116 Church St. SE, Minneapolis, MN 55455 USA}
\doi{}

\author{N. Gupta}
\affiliation{Australia Telescope National Facility, CSIRO, Space and Astronomy, PO Box 1130, Bentley, WA 6151, Australia}

\author{E. Carretti}
\affiliation{INAF - Istituto di Radioastronomia, Via Gobetti 101, 40129 Bologna, Italy}

\author{C.~S.~Anderson}
\affiliation{Research School of Astronomy \& Astrophysics, Australian National University, Canberra, ACT 2611, Australia}

\author{S. Dai}
\affiliation{Australia Telescope National Facility, CSIRO, Space and Astronomy, P.O. Box 76, Epping, NSW 1710, Australia}

\author{G. G\"urkan}
\affiliation{Centre for Astrophysics Research, University of Hertfordshire, College Lane, Hatfield AL10 9AB, UK}
\alsoaffiliation{Australia Telescope National Facility, CSIRO Space and Astronomy, PO Box 1130, Bentley WA 6102, Australia}

\author{D. Parkinson}
\affiliation{Korea Astronomy and Space Science Institute, 776 Daedeok-daero, Yuseong-gu, Daejeon 34055, Republic of Korea}

\author{I. Prandoni}
\affiliation{INAF - Istituto di Radioastronomia, Via Gobetti 101, 40129 Bologna, Italy}

\author{S. Riggi}
\affiliation{INAF - Osservatorio Astrofisico di Catania, Via Santa Sofia 78, I-95123 Catania, Italy}

\author{C.~S. Saraf}
\affiliation{Korea Astronomy and Space Science Institute, 776 Daedeok-daero, Yuseong-gu, Daejeon 34055, Republic of Korea}

\author{Y.~K.~Ma}
\affiliation{Research School of Astronomy \& Astrophysics, Australian National University, Canberra, ACT 2611, Australia}

\author{M.~D.~Filipovi\'c}
\affiliation{Western Sydney University, Locked Bag 1797, Penrith South DC, NSW 2751, Australia}

\author{G. Umana}
\affiliation{INAF - Osservatorio Astrofisico di Catania, Via Santa Sofia 78, I-95123 Catania, Italy}

\author{B.~Bahr-Kalus}
\affiliation{INAF - Osservatorio Astrofisico di Torino, Via Osservatorio 20, I-10025 Pino Torinese, Italy}
\alsoaffiliation{INFN - Sezione di Torino, Via P. Giuria 1, 10125 Torino, Italy}
\alsoaffiliation{Dipartimento di Fisica, Universit\`a degli Studi di Torino, Via P.\ Giuria 1, 10125 Torino, Italy}

\author{B.~S.~Koribalski}
\affiliation{Australia Telescope National Facility, CSIRO, Space and Astronomy, P.O. Box 76, Epping, NSW 1710, Australia}
\alsoaffiliation{Western Sydney University, Locked Bag 1797, Penrith South DC, NSW 2751, Australia}

\author{E.~Lenc}
\affiliation{Australia Telescope National Facility, CSIRO, Space and Astronomy, P.O. Box 76, Epping, NSW 1710, Australia}

\author{A. Ingallinera}
\affiliation{INAF - Osservatorio Astrofisico di Catania, Via Santa Sofia 78, I-95123 Catania, Italy}

\author{J. Afonso}
\affiliation{Instituto de Astrof\'{i}sica e Ci\^{e}ncias do Espa\c{c}o, Universidade de Lisboa, OAL, Tapada da Ajuda, 1349-018 Lisbon, Portugal}
\alsoaffiliation{Departamento de F\'{i}sica, Faculdade de Ci\^{e}ncias, Universidade de Lisboa, Edif\'{i}cio C8, Campo Grande, 1749-016 Lisbon, Portugal}

\author{A. Ahmad}
\affiliation{Western Sydney University, Locked Bag 1797, Penrith South DC, NSW 2751, Australia}

\author{U.~T. Ahmed}
\affiliation{Australian Astronomical Optics, Macquarie University, 105 Delhi Rd, North Ryde, NSW 2113, Australia}

\author{E. L. Alexander}
\affiliation{School of Physics \& Astronomy, University of Leeds, Leeds, LS2 9JT, UK}

\author{H. Andernach}
\affiliation{Th\"uringer Landessternwarte, Sternwarte 5,
D-07778 Tautenburg, Germany}
\alsoaffiliation{Departamento de Astronom\'ia, Universidad de Guanajuato,
DCNE, Callej\'on de Jalisco s/n, Guanajuato, C.P. 36023, GTO, Mexico} 

\author{J. Asorey}
\affiliation{Departamento de F\'isica Te\'orica, Centro de Astropart\'iculas y F\'isica de Altas Energ\'ias, Universidad de Zaragoza, 50009 Zaragoza, Spain}

\author{A. J. Battisti}
\affiliation{International Centre for Radio Astronomy Research (ICRAR), University of Western Australia, M468, 35 Stirling Highway, Crawley, WA 6009, Australia}
\alsoaffiliation{Research School of Astronomy \& Astrophysics, Australian National University, Canberra, ACT 2611, Australia}
\alsoaffiliation{ARC Centre of Excellence for All Sky Astrophysics in 3 Dimensions (ASTRO 3D), Australia}

\author{M. Bilicki}
\affiliation{Center for Theoretical Physics, Polish Academy of Sciences, al. Lotnik\'{o}w 32/46, 02-668 Warsaw, Poland}

\author{A. Botteon}
\affiliation{INAF - Istituto di Radioastronomia, Via Gobetti 101, 40129 Bologna, Italy}

\author{M. J. I. Brown}
\affiliation{School of Physics \& Astronomy, Monash University, Clayton, VIC 3800, Australia}

\author{M. Br\"uggen}
\affiliation{Hamburger Sternwarte, University of Hamburg, Gojenbergsweg 112, 21029 Hamburg, Germany}

\author{M. Cowley}
\affiliation{School of Chemistry \& Physics, Faculty of Science, Queensland University of Technology, Brisbane, QLD 4000, Australia}
\alsoaffiliation{University of Southern Queensland, Centre for Astrophysics, West Street, Toowoomba QLD 4350, Australia}

\author{K. C. Dage}
\affiliation{International Centre for Radio Astronomy Research -- Curtin University, GPO Box U1987, Perth, WA 6845, Australia}

\author{C. L. Hale}
\affiliation{Astrophysics, Denys Wilkinson Building, Department of Physics, University of Oxford, Keble Road, OX1 3RH, UK}

\author{M.J. Hardcastle}
\affiliation{Centre for Astrophysics Research, University of Hertfordshire, College Lane, Hatfield AL10 9AB, UK}

\author{R. Kothes}
\affiliation{Dominion Radio Astrophysical Observatory, Herzberg Research Centre for Astronomy and Astrophysics, National Research Council Canada, PO Box 248, Penticton, BC V2A 6J9, Canada}
\alsoaffiliation{Department of Physics, University of Alberta, Edmonton, Alberta, T6G 2E1, Canada}

\author{S. Lazarevi\'c}
\affiliation{Western Sydney University, Locked Bag 1797, Penrith South DC, NSW 2751, Australia}
\alsoaffiliation{Australia Telescope National Facility, CSIRO, Space and Astronomy, P.O. Box 76, Epping, NSW 1710, Australia}
\alsoaffiliation{Astronomical Observatory, Volgina 7, 11060 Belgrade, Serbia}

\author{Y.-T. Lin}
\affiliation{Institute of Astronomy and Astrophysics, Academia Sinica (ASIAA), Taipei, 10617, Taiwan}

\author{K. J. Luken}
\affiliation{Western Sydney University, Locked Bag 1797, Penrith South DC, NSW 2751, Australia}

\author{J.~P.~Moss}
\affiliation{School of Chemical and Physical Sciences, Victoria University of Wellington, Kelburn, Wellington 6012}

\author{J. Prathap}
\affiliation{School of Mathematical and Physical Sciences, 12 Wally's Walk, Macquarie University, NSW 2109, Australia}
\alsoaffiliation{Macquarie University Astrophysics and Space Technologies Research Centre, Sydney, NSW 2109, Australia}
\alsoaffiliation{ARC Centre of Excellence for All Sky Astrophysics in 3 Dimensions (ASTRO 3D), Australia}

\author{S. F. Rahman}
\affiliation{Lahore University of Management Sciences (LUMS), Pakistan}
\alsoaffiliation{NCBC at NED University of Engineering and Technology, Pakistan}

\author{T.~H. Reiprich}
\affiliation{Argelander Institute for Astronomy, University of Bonn, Auf dem H\"ugel 71, 53121 Bonn, Germany}

\author{C. J. Riseley}
\affiliation{Astronomisches Institut der Ruhr-Universit\"{a}t Bochum (AIRUB), Universit\"{a}tsstra{\ss}e 150, 44801 Bochum, Germany}
\alsoaffiliation{Dipartimento di Fisica e Astronomia, Universit\`a degli Studi di Bologna, via P. Gobetti 93/2, 40129 Bologna, Italy}
\alsoaffiliation{INAF -- Istituto di Radioastronomia, via P. Gobetti 101, 40129 Bologna, Italy}

\author{M. Salvato}
\affiliation{Max-Planck Institute for Extraterrestrial Physics, Giessenbachstrasse 1, 85748 Garching, Germany}

\author{N. Seymour}
\affiliation{International Centre for Radio Astronomy Research, Curtin University, GPO Box U1987, Bentley, WA 6845, Australia}

\author{S. S. Shabala}
\affiliation{School of Natural Sciences, University of Tasmania, Private Box 37, Hobart, TAS 7001, Australia}

\author{D. J. B. Smith}
\affiliation{Centre for Astrophysics Research, University of Hertfordshire, College Lane, Hatfield AL10 9AB, UK}

\author{M. Vaccari}
\affiliation{Inter-University Institute for Data Intensive Astronomy, and Department of Physics and Astronomy, University of the Western Cape, Robert Sobukwe Road, 7535 Bellville, Cape Town, South Africa}
\alsoaffiliation{Inter-University Institute for Data Intensive Astronomy, Department of Astronomy, University of Cape Town, 7701 Rondebosch, Cape Town, South Africa}
\alsoaffiliation{INAF - Istituto di Radioastronomia, via Gobetti 101, 40129 Bologna, Italy}

\author{J. Th. van Loon}
\affiliation{Lennard-Jones Laboratories, Keele University, ST5 5BG, UK}

\author{O. I. Wong}
\affiliation{Australia Telescope National Facility, CSIRO, Space and Astronomy, PO Box 1130, Bentley, WA 6151, Australia}
\alsoaffiliation{ICRAR, The University of Western Australia, 35 Stirling Hwy, 6009 Crawley, Australia}

\author{R. Z. E. Alsaberi}
\affiliation{Faculty of Engineering, Gifu University, 1-1 Yanagido, Gifu 501-1193, Japan}
\alsoaffiliation{Western Sydney University, Locked Bag 1797, Penrith South DC, NSW 2751, Australia}

\author{A. D. Asher}
\affiliation{Western Sydney University, Locked Bag 1797, Penrith South DC, NSW 2751, Australia}
\alsoaffiliation{Australia Telescope National Facility, CSIRO, Space and Astronomy, P.O. Box 76, Epping, NSW 1710, Australia}

\author{B. D. Ball}
\affiliation{Department of Physics, University of Alberta, Edmonton, Alberta, T6G 2E1, Canada}

\author{D. Barbosa}
\affiliation{Instituto de Astrof\'{i}sica e Ci\^{e}ncias do Espa\c{c}o, Universidade de Lisboa, OAL, Tapada da Ajuda, 1349-018 Lisbon, Portugal}
\alsoaffiliation{Departamento de F\'{i}sica, Faculdade de Ci\^{e}ncias, Universidade de Lisboa, Edif\'{i}cio C8, Campo Grande, 1749-016 Lisbon, Portugal}

\author{N. Biava}
\affiliation{Th\"uringer Landessternwarte, Sternwarte 5, D-07778 Tautenburg, Germany}
\alsoaffiliation{INAF - IRA, via P. Gobetti 101, I-40129 Bologna, Italy}

\author{A. C. Bradley}
\affiliation{Western Sydney University, Locked Bag 1797, Penrith South DC, NSW 2751, Australia}

\author{R. Carvajal}
\affiliation{Instituto de Astrof\'{i}sica e Ci\^{e}ncias do Espa\c{c}o, Universidade de Lisboa, OAL, Tapada da Ajuda, 1349-018 Lisbon, Portugal}
\alsoaffiliation{Departamento de F\'{i}sica, Faculdade de Ci\^{e}ncias, Universidade de Lisboa, Edif\'{i}cio C8, Campo Grande, 1749-016 Lisbon, Portugal}

\author{E. J. Crawford}
\affiliation{Western Sydney University, Locked Bag 1797, Penrith South DC, NSW 2751, Australia}

\author{T. J. Galvin}
\affiliation{Australia Telescope National Facility, CSIRO, Space and Astronomy, PO Box 1130, Bentley, WA 6151, Australia}

\author{M. T. Huynh}
\affiliation{Australia Telescope National Facility, CSIRO, Space and Astronomy, PO Box 1130, Bentley, WA 6151, Australia}
\alsoaffiliation{ICRAR, The University of Western Australia, 35 Stirling Hwy, 6009 Crawley, Australia}

\author{D. A. Leahy}
\affiliation{Department of Physics and Astronomy, University of Calgary, Calgary, AB, T2N 1N4, Canada}

\author{I. Matute}
\affiliation{Instituto de Astrof\'{i}sica e Ci\^{e}ncias do Espa\c{c}o, Universidade de Lisboa, OAL, Tapada da Ajuda, 1349-018 Lisbon, Portugal}
\alsoaffiliation{Departamento de F\'{i}sica, Faculdade de Ci\^{e}ncias, Universidade de Lisboa, Edif\'{i}cio C8, Campo Grande, 1749-016 Lisbon, Portugal}

\author{V. A. Moss}
\affiliation{Australia Telescope National Facility, CSIRO, Space and Astronomy, P.O. Box 76, Epping, NSW 1710, Australia}

\author{C. Pappalardo}
\affiliation{Instituto de Astrof\'{i}sica e Ci\^{e}ncias do Espa\c{c}o, Universidade de Lisboa, OAL, Tapada da Ajuda, 1349-018 Lisbon, Portugal}
\alsoaffiliation{Departamento de F\'{i}sica, Faculdade de Ci\^{e}ncias, Universidade de Lisboa, Edif\'{i}cio C8, Campo Grande, 1749-016 Lisbon, Portugal}

\author{Z. J. Smeaton}
\affiliation{Western Sydney University, Locked Bag 1797, Penrith South DC, NSW 2751, Australia}

\author{V. Velovi\'c}
\affiliation{Western Sydney University, Locked Bag 1797, Penrith South DC, NSW 2751, Australia}

\author{T. Zafar}
\affiliation{School of Mathematical and Physical Sciences, 12 Wally's Walk, Macquarie University, NSW 2109, Australia}

\received {dd Mmm YYYY}
\revised  {dd Mmm YYYY}
\accepted {dd Mmm YYYY}
\published{22 September 2020}


\begin{document}
\maketitle

\begin{abstract}
We present the Evolutionary Map of the Universe (EMU) survey conducted with the Australian Square Kilometre Array Pathfinder (ASKAP). EMU aims to deliver the touchstone radio atlas of the southern hemisphere. We introduce EMU and review its science drivers and key science goals, updated and tailored to the current ASKAP five-year survey plan. The development of the survey strategy and planned sky coverage is presented, along with the operational aspects of the survey and associated data analysis, together with a selection of diagnostics demonstrating the imaging quality and data characteristics. We give a general description of the value-added data pipeline and data products before concluding with a discussion of links to other surveys and projects and an outline of EMU's legacy value.
\end{abstract}

\section{INTRODUCTION }
\label{sec:int}
\subsection{EMU and radio surveys}
The ``Evolutionary Map of the Universe'' \citep[EMU,][]{2011PASA...28..215N,2021PASA...38...46N}\footnote{ EMU Project page: \url{https://emu-survey.org/}} is a landmark project to deliver the touchstone radio atlas of the southern hemisphere sky using the Australian Square Kilometre Array Pathfinder (ASKAP) radio telescope \citep{2021PASA...38....9H}. 

There is a long history of pushing radio telescopes to their limits to maximise sky coverage at the best possible sensitivity and resolution to understand the nature and properties of ever fainter radio source populations \citep[e.g.,][]{1976A&AS...25..453W,1985A&A...146...38K,1987A&AS...71..221O,1993ApJ...405..498W,1998MNRAS.296..839H,1999MNRAS.305..297G,2000A&AS..146...31P,2002AJ....123.1784D,2003AJ....125..465H,2004AJ....128.1974S,2005AJ....130.1373H,2006MNRAS.372..741S,2006AJ....132.2409N,2007MNRAS.378..995M,2017A&A...602A...1S,2021A&A...648A...2S,Best2023, Hale2024}. These efforts have revealed that the bright (flux densities above a few mJy) extragalactic radio source population is composed primarily of systems powered by supermassive black holes \citep[e.g.,][]{2020PASA...37...18W,2020PASA...37...17W}, both nearby and extending to the highest redshifts, while the fainter sources are dominated by star-forming galaxies and low luminosity or ``radio quiet'' (RQ) active galactic nuclei (AGN) systems \citep[e.g.,][]{2008MNRAS.386.1695S,2015MNRAS.448.2665W,2017MNRAS.468..217W,2018MNRAS.481.4548P,2021MNRAS.506.3540P,Drake2024}. Wide area sky surveys \citep[e.g.,][]{1995ApJ...450..559B,1998AJ....115.1693C,2003MNRAS.342.1117M,
2017A&A...598A..78I,2017MNRAS.464.1146H,2022PASA...39...35H,2022A&A...659A...1S} have illustrated the complexity of radio emission associated with radio galaxy jets and lobes \citep[e.g.,][]{2022MNRAS.512.6104G,2024MNRAS.533..608K}, large scale structures in galaxy clusters \citep[e.g.,][]{giovannini99, kempner01, duchesne21eor, Duchesne2024, botteon22}, and from supernova remnants (SNRs), neutral atomic hydrogen (H\,{\sc i}) emission, and other objects and structures in the Galactic Plane \citep[e.g.,][]{2021MNRAS.506.2232U, 2023AJ....166..149F,2024PASA...41...32L, 2024MNRAS.534.2918S} and Magellanic Clouds \citep[e.g.,][]{2021MNRAS.506.3540P}.

Each new generation of radio telescope technology and improved radio survey scale has enhanced our understanding of the Universe. These developments continued in the late 2000s and early 2010s with extensive preparation worldwide for major projects anticipating the advent of Square Kilometre Array (SKA) precursor facilities \citep{2013PASA...30...20N}, and ultimately the SKA itself. In the northern hemisphere, the Low-frequency Array (LOFAR), has pushed the limits at low frequencies \citep{2021A&A...648A...2S,2022A&A...659A...1S, 2022NatAs...6..350S,deGasperin2023, groeneveld24, 2024A&A...689A..80D}. A key driver for these was the capability of such new facilities to move beyond the practical limitations of then-existing telescopes \citep[e.g.,][]{2013PASA...30...20N,2017NatAs...1..671N}. In addition to the many planned scientific developments that such major projects could achieve, it has long been established that expanding the available observational parameter space in this fashion leads to new discoveries beyond just those that can be foreseen \citep{2009arad.workE...7E,2017PASA...34....7N}.
It was in this environment and with this sense of excitement and anticipation that the original concept for the EMU project was formed.

\subsection{EMU history}
EMU was conceived in 2009, crystallising earlier ideas around the concept of a maximal area highly-sensitive radio continuum science project with the ASKAP radio telescope \citep{2007PASA...24..174J,2008ExA....22..151J}. EMU was one of two concepts equally ranked in that year as the highest priority projects that ASKAP should deliver, the other being the Widefield ASKAP L-band Legacy All-sky Blind surveY \citep[WALLABY,][]{2020Ap&SS.365..118K} which is focussed on detecting neutral hydrogen spectral line emission (H\,{\sc i}) in the nearby Universe but also delivers deep 1.4\,GHz radio continuum data \citep{Koribalski2012,2020Ap&SS.365..118K}. Closely linked was the Polarisation Sky Survey of the Universe's Magnetism (POSSUM) survey \citep[][Gaensler et al., 2025, in press]{2010AAS...21547013G,2017ASPC..512..359P} to measure the continuum source polarisation properties and Faraday rotation, to develop the best insights into the role of magnetic fields in the Universe. EMU, WALLABY, and POSSUM together were originally conceived as complementary projects that would be carried out commensally through a single observing program with ASKAP. They each capitalise on ASKAP's unique phased-array feed receiver technology \citep{2009IEEEP..97.1507D,2015icea.conf..541C,2021PASA...38....9H} that allows for a very rapid survey speed, a key development necessary for delivering very sensitive all-hemisphere programs. These surveys will provide targets for the SKA, which will not conduct such all-sky surveys itself.

The original EMU concept, detailed in \citet{2011PASA...28..215N}, was for a $3\pi\,$sr sky survey from the South Celestial Pole up to $\delta = +30^{\circ}$ reaching to a root-mean-square (rms) noise level of $\sigma=10\,\mu$Jy\,beam$^{-1}$ with a resolution of $\sim 10''$. Such coverage and sensitivity could deliver a survey cataloguing as many as 70 million radio sources at a frequency of 1.4\,GHz. As ASKAP commissioning progressed in the late 2010s it became clear that a more realistic performance goal would be a $3\pi$\,sr sky survey conducted at a frequency around 900\,MHz, and limited to a sensitivity of $\sigma=20-30\,\mu$Jy\,beam$^{-1}$ with a resolution of $15''$, a consequence arising from a combination of telescope technical performance, the radio frequency interference (RFI) environment at the telescope site \citep{lourencco2024survey}, and practical observation scheduling reasons. The original plan for a fully commensal observing program for EMU and WALLABY also became impractical due to the RFI. Mapping H\,{\sc i} in the nearby Universe requires WALLABY to observe close to 1.4\,GHz, while EMU, in order to retain the best sensitivity and survey speed, moved to a lower frequency where ASKAP's continuum sensitivity is optimal. In parallel, the POSSUM project established a clear desire for commensal observing and data sharing with either or (ideally) both EMU and WALLABY.

\begin{figure*}[hbt!]
\centering
\includegraphics[width=0.9\textwidth]{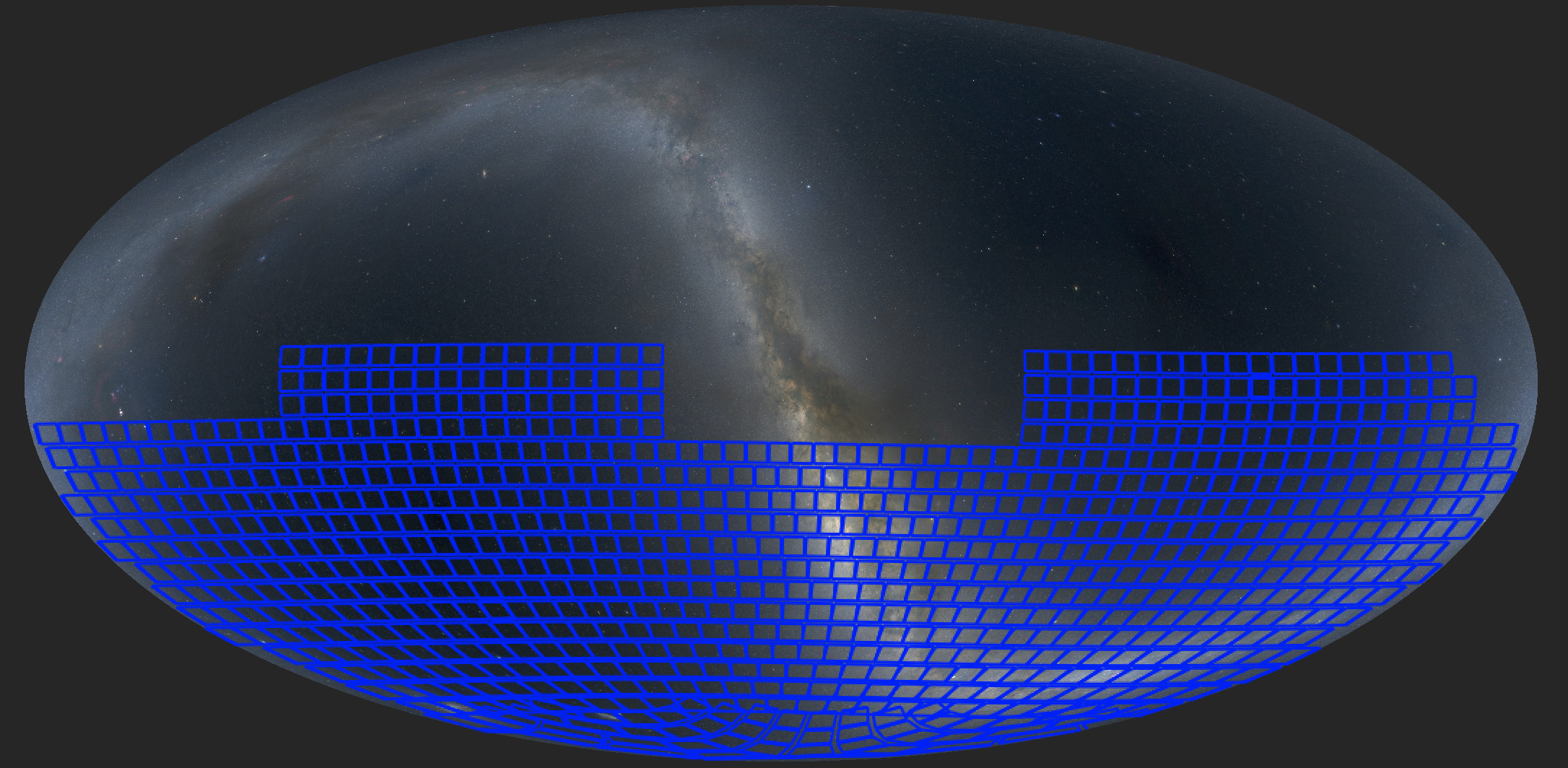}
\caption{The EMU sky coverage to be delivered in 2028. The background image is the ``Mellinger coloured'' image \citep{2009PASP..121.1180M} accessed through AladinLite \citep{2014ascl.soft02005B}. Each blue outline represents the footprint of a single ASKAP tile, and there are 853 such footprints comprising the full EMU survey. There is a small overlap between each adjacent tile. North of $\delta=-10^{\circ}$ each footprint requires two observations, leading to the total of 1014 tile observations (see \S\,\ref{sec:survey}). North of $\delta=-70^{\circ}$, the tiles follow constant declination strips. Further south, to efficiently cover the pole, the tiles are arrayed in a rectilinear grid centred on the pole.}
\label{fig:coverage}
\end{figure*}

A review in 2021 of the ASKAP survey science projects, while the telescope was finalising its commissioning activities, reinforced the strong rankings of EMU (jointly with POSSUM) and WALLABY, and recommended that EMU be awarded a total of 8533\,h of ASKAP observing time over the five-year strategic timeline being considered (originally 2022-2027). In line with this allocation, the original $3\pi\,$sr survey goal was reduced to a coverage of $2\pi\,$sr, although the original goal remains as a future ambition for EMU following the initial five-year operational period of ASKAP.
With these modified survey goals, EMU now anticipates cataloguing about 20 million extragalactic radio sources. This prediction is derived from the Tiered Radio Extragalactic Continuum Simulation \citep[T-RECS;][]{2019MNRAS.482....2B}, a simulation
of the radio continuum properties of the two main extragalactic radio populations (AGN and star-forming galaxies) over the 150\,MHz to 20\,GHz range. The final EMU survey five-year coverage is shown in Figure~\ref{fig:coverage}.

\subsection{Current EMU status}
ASKAP initiated formal full survey operations in May 2023. EMU observations are projected to be complete in 2028. As of the date of writing (April 2025), out of 1014 total tiles (see details in \S\,\ref{sec:obs}), there are 307 ($30\,$\%) that have been validated as good
and released through the CSIRO ASKAP Science Data Archive (CASDA)\footnote{https://research.csiro.au/casda/} (see details in \S\,\ref{sec:obs}). 

This paper details the scientific motivations for conducting EMU (\S\,\ref{sec:sci}), the EMU survey design (\S\,\ref{sec:survey}), the observations, data processing and validation (\S\,\ref{sec:obs}), EMU source statistics (\S\,\ref{sec:cataloguestats}), and an overview of the value-added data pipeline (\S\,\ref{sec:data}).
This is followed by a discussion of EMU in the wider context, including related survey programs (\S\,\ref{sec:discussion}). We summarise and present the next steps for EMU in \S\,\ref{sec:conclusion}.

Throughout, where relevant, we assume cosmological parameters of $H_0=70\,$km\,s$^{-1}$\,Mpc$^{-1}$, $\Omega_M=0.3$,
$\Omega_\Lambda=0.7$ and $\Omega_{\rm{k}} = 0$.

\section{SCIENTIFIC GOALS}
\label{sec:sci}
EMU science spans a vast range of astrophysics and cosmology, and has remained broadly the same as in the original EMU concept \citep{2011PASA...28..215N}, albeit with some evolution as the fields have progressed over the past decade. Here we review the key scientific areas that EMU is primarily aimed at addressing, while also acknowledging there will be a vast wealth of science supported by the survey that extends well beyond these goals. This extends to an expectation of many legacy science outcomes not even anticipated at this early stage \citep[e.g.,][]{2017PASA...34....7N}.

\subsection{Star-forming galaxies and AGN}
EMU will detect about 20 million sources to a $5\sigma$ limit of $100\,\mu$Jy\,beam$^{-1}$. Of these, about $4-5$ million will have radio emission dominated by AGN, while $15-16$ million will have emission dominated by star formation (SF), based on predictions from the T-RECS simulations \citep{2019MNRAS.482....2B}. 
Both populations will span a significant fraction of the age of the Universe, up to reionisation for radio AGN and the most extreme starbursts. EMU will allow large-scale statistical exploration of the evolution of these populations and how it depends on galaxy mass, environment, SF history, interactions and merger history \citep[e.g.,][]{2004ApJ...615..209H,2008MNRAS.386.1695S,2017MNRAS.466.2312D,2017A&A...602A...5N}. Massive galaxies appear to form their stars early and quickly, progressively becoming less active after redshift $z\sim 2$, while lower-mass galaxies become dominant at lower redshifts \citep[e.g.,][]{2009ApJ...690.1074M}. This evolution is mirrored in the AGN accretion rate, suggesting some feedback mechanism couples AGN to galaxy evolution \citep[e.g.,][]{2016MNRAS.457..629C,2023ApJ...959L..18D}.
EMU will quantify these effects in detail, by providing a deep homogeneously selected sample of both AGN and SF galaxies over the majority of cosmic history, unbiased by dust obscuration \citep[e.g.,][]{2003ApJ...597..269A}.

\subsubsection{The Obscured Universe}
\label{sec:dust}
In the context of dust obscuration, radio observations such as EMU have a critical role to play in uncovering the highly obscured Universe, in a way that complements optical and ultraviolet (UV) surveys. It has been long established that radio-selected samples contain more heavily obscured systems than optically-selected samples \citep[e.g.,][]{2003ApJ...597..269A}. But importantly, a recent EMU analysis \citep{2024PASA...41...21A} demonstrates that even in an optically-selected parent sample, radio-detected galaxies exhibit significantly higher levels of dust obscuration, especially for low-mass, low star formation rate (SFR) galaxies.
Such results suggest that a substantial fraction of the cosmic SFR density and black hole accretion history may be hidden in optically obscured systems. Through EMU’s unparalleled sensitivity over its extensive survey area, we can systematically characterise the dust-obscured universe. This will allow us to investigate how the prevalence and properties of such galaxies evolve with redshift, galaxy mass, and local environment.

\subsubsection{The Cosmic Star Formation History}
Star forming galaxies identified through EMU will provide an unprecedented view of the cosmic star formation history \citep[CSFH; e.g.,][]{2004ApJ...615..209H,2006ApJ...651..142H,2017A&A...602A...5N, 2018MNRAS.475.2891D, 2023MNRAS.523.6082C} and its link to the cosmic stellar mass density history \citep[CSMH; e.g.,][]{2008MNRAS.391..363W,2008MNRAS.385..687W}. This relies on both accurate AGN classification \citep[e.g.,][]{2010MNRAS.403.1036C,2014ARA&A..52..589H}, and robust SFR estimation \citep[e.g.,][]{2016MNRAS.461..458D, 2017MNRAS.466.2312D, 2017ApJ...847..136B}.

To develop improved radio-based SFR calibrations, EMU’s data will be combined with multi-wavelength photometric and spectroscopic data from surveys such as the Galaxy and Mass Assembly \citep[GAMA,][]{2011MNRAS.413..971D, 2022MNRAS.513..439D} and the Wide-Area Vista Extragalactic Survey \citep[WAVES,][]{2019Msngr.175...46D}. Drawing on the latest developments in population synthesis tools, \citep[e.g.,][]{2024arXiv241017697R,2024arXiv241017698B}, AGN contributions can be explicitly accounted for, and radio photometry can be used to refine SFR estimates. In turn, samples with extensive multiwavelength data can be used to inform and improve the calibration of radio luminosities to SFRs for those systems without such extensive supporting measurements.
Given EMU's very large sample sizes, different populations can then be separated by astrophysically relevant quantities, such as stellar mass and environment, to construct the CSFH and CSMH for each subset, to explore the mass and environment dependence of the growth of stellar mass in galaxies, along with its link to AGN. The joint constraint of the CSFH and CSMH can also be used to investigate the cosmic evolution of the stellar initial mass function \citep{2008MNRAS.385..687W,2018PASA...35...39H} separated by such populations.

In cases where radio sources lack counterparts, machine learning \citep[ML; e.g.,][]{2023PASA...40...39L} or statistical techniques (e.g., Prathap et al., submitted) can be applied to assign redshifts and to infer SFRs and stellar masses probabilistically, enabling population-based analyses even for radio sources without counterparts. This comprehensive effort will involve significant development work on SFR calibrations, AGN/SF diagnostic techniques, ML applications, and redshift and stellar mass assignments, further advancing these fields in addition to our understanding of the cosmic SF and mass history through EMU.

\subsubsection{The AGN and Star Formation Link}
\label{sec:feedback}
Construction of obscuration independent samples of AGN hosts is vital in order to robustly establish the AGN duty cycle and its links to SF, the relative timing of AGN and SF activity in galaxies exhibiting both phenomena \citep{2007MNRAS.382.1415S, 2010MNRAS.405..933W, 2012MNRAS.423...59S, 2017MNRAS.464.4706S}, as well as to galaxy transitions through post-starburst or ``green valley'' stages \citep[e.g.,][]{2022MNRAS.515.6046P}, allowing for a comprehensive overview of galaxy evolution. Complementary multiwavelength photometry and redshifts for all elements of these analyses are critical, not only to supplement EMU detections but also to identify AGN hosts that are not dominated by radio emission, such as through X-rays and infrared. Consequently, maximising the EMU survey area is necessary to encompass key complementary surveys in different areas of the sky. It is equally critical to maximise the sample numbers, especially at high-$z$, due to the inevitable reduction in sample size necessary when measuring evolutionary effects. This arises from (1)~limited numbers of counterparts identified in complementary surveys, and (2)~construction of luminosity- or mass-limited subsamples split by redshift, mass, environment, galaxy type, and more. This is compounded when the necessary complementary data (different for different types of analyses) only exist over limited regions of sky.

A crucial piece of the galaxy evolution puzzle is the role played by AGN in regulating SF in the host galaxy, commonly referred to as AGN feedback \citep{2006MNRAS.365...11C}.
While the NRAO VLA sky survey \citep[NVSS,][]{condon1998nrao}, the Faint Images of the Radio Sky at Twenty centimeters \citep[FIRST,][]{becker1995first} and the Rapid ASKAP Continuum Survey \citep[RACS,][]{2020PASA...37...48M,Hale2021} are largely dominated by bright radio galaxy populations, EMU is ideal for studying the low luminosity tail of radio-loud (RL) AGN, as well as the radio-quiet (RQ) AGN population, which become significant at S$_\mathrm{1.4 \; GHz}<100-200 \; \mu$Jy \citep{2013MNRAS.436.3759B}.
RQ AGN show signatures of nuclear activity at optical, infrared or X-ray bands \citep[e.g.][]{Best2023,Das2024,Drake2024}, but are comparatively faint in the radio domain. To identify the RQ AGN population, EMU can rely on spectroscopy from GAMA, Sloan Digital Sky Survey \citep[SDSS,][]{2019BAAS...51g.274K}, Dark Energy Spectroscopic Instrument \citep[DESI,][]{aghamousa2016desi}, the William Herschel Telescope Enhanced Area Velocity Explorer instrument \citep[WEAVE,][]{2012SPIE.8446E..0PD} and soon also 4MOST surveys including WAVES, 4MOST Hemisphere Survey \citep[4HS,][]{2023Msngr.190...46T} and the Optical Radio Continuum and H\,{\sc i} Deep Spectroscopic Survey \citep[ORCHIDSS,][]{Duncan2023}, as well as WEAVE-LOFAR \citep{Smith2016} and \textit{Euclid} \citep{scaramella2022euclid}. Photometry in the optical, infrared, and X-ray will also be obtained from telescopes and surveys such as \textit{Gaia} \citep{Gaia2016, Gaia2023}, the Dark Energy Survey \citep[DES,][]{2016MNRAS.460.1270D}, \textit{Herschel} \citep{pilbratt2010herschel} and \textit{eROSITA} \citep{predehl2021erosita} respectively.

This is illustrated by extensive work drawing on the EMU Early Science observations of the GAMA 23 region, covering an 80 deg$^{2}$ area \citep{2022MNRAS.512.6104G}, which links EMU and GAMA data. This resource has already been used extensively to explore aspects of AGN \citep{2024PASA...41...16P}, SF (Ahmed et al., submitted), and dust obscuration in galaxies \citep[][]{2024PASA...41...21A}. \citet{2022MNRAS.512.6104G} used the GAMA spectroscopy to identify RQ AGN, marking the first significant work in this domain. Candini et al. (in prep) are expanding on earlier work \citep{2013MNRAS.433..622M} using SDSS and NVSS that show that bright ($L$[O\,{\sc iii}]$ \; >10^{42}$ erg s$^{-1}$) AGN characterised by radio luminosities ($L[1.4 \; \rm{GHz}]>10^{23}$ W Hz$^{-1}$) tend to have larger [O\,{\sc iii}] line width. This suggests that [O\,{\sc iii}] outflows may be linked to the presence of a radio AGN in bright systems, consistent with the increased fraction of high excitation radio galaxies (HERGs) at the highest radio luminosities \citep{2012MNRAS.421.1569B}. Despite their relative rarity, these bright radio galaxies dominate the kinetic feedback budget from AGN \citep{2015ApJ...806...59T, 2019A&A...622A..12H}.

Radio observations of AGN at low redshift are extensive compared to the scarcity at high redshift.
Only a handful are known at $z>6$, with the highest-redshift radio source at $z=6.8$ \citep{2021ApJ...909...80B} and with the {\em James Webb Space Telescope\/} (\textit{JWST}) finding a strong candidate at $z\sim 7.7$ \citep{lambrides:24}. Many more high-redshift AGN are likely to be among the EMU sources \citep{2023MNRAS.519.4902S}, but cannot yet be identified because their redshifts have not been measured. Indeed, RACS \citep{2020PASA...37...48M, Hale2021} has already found detections of new RL quasi-stellar objects (QSOs) at $z>6$ \cite{ighina:21}.
AGN near the end of cosmic reionisation contain the (progenitors of the) highest-mass black holes and are the youngest radio sources in the Universe. High-$z$ AGN are critical in understanding the growth of black holes in the early Universe (through rapid accretion), feedback on host galaxies, and their luminosity function and spatial distribution are important for cosmological studies. While extreme dust obscuration and neutral hydrogen (H\,{\sc i}) absorption in such systems pose a significant challenge for optical, infrared and X-ray observations, they are transparent to radio emission. Radio observations can detect dust-obscured young AGN in the transition phase of galaxy evolution (from post-merger to quasar), and are a necessary probe of high-redshift galactic environment. A key question is where the radio emission of these high-redshift infrared-luminous sources arises. It may be attributed to a weak jet, quasar winds, disk winds, nuclear starbursts, or something else \citep[e.g.][]{2019NatAs...3..387P}. Because of the relatively low radio luminosity and surface density on the sky of these high-$z$ sources, highly sensitive and wide-area observations, like those provided by surveys such as EMU and LOFAR, are needed. Low frequency southern hemisphere radio data are also available from the Murchison Wide-field Array (MWA) surveys GaLactic and Extragalactic All-sky MWA survey \citep[GLEAM,][]{2017MNRAS.464.1146H} and GLEAM-eXtended \citep[GLEAM-X,][]{2022PASA...39...35H,2024PASA...41...54R}, which, despite having much poorer spatial resolution than EMU, will still be important resources for constraining radio spectral indices for the brighter EMU sources.

\subsection{Astrophysics of radio galaxies}
By ``radio galaxies'' here we are referring to those systems with radio jets and lobes associated with an AGN.
With EMU's extremely good surface brightness sensitivity \citep[e.g.,][]{2021A&A...647A...3B,2021PASA...38....3N}, well-resolved extended emission from such radio galaxies will be detected for several $10^5$ sources, based on extrapolations from the numbers found in the first EMU Pilot Survey \citep{2021PASA...38...46N}. Such large samples of radio galaxies will change our understanding of their overall structure. Detailed studies of their extended structure with EMU \citep{2022MNRAS.516.1865V} will influence models of the powering jets as well as the interactions with the surrounding medium \citep[e.g., ][]{2016MNRAS.461.2025E,2021MNRAS.508.5239Y,2023PASA...40...14Y}. In addition, unusual structures which challenge our models are important but rare \citep[e.g.][]{2024MNRAS.533..608K, 2024MNRAS.532.3682K}, and can only be found by surveying sufficiently large areas of sky with superb surface brightness sensitivity. Spectral index maps for such extended objects can be derived from EMU data, and will be available in unprecedented numbers, enabling the history of relativistic particle gains and losses in radio galaxies to be systematically studied.

EMU's excellent surface brightness sensitivity ensures efficient detection of giant radio galaxies, including those of relatively low luminosities. 
Giant radio galaxies are examples of both extreme jet physics and low density environments, making them a valuable probe of astrophysics in extreme conditions. Expanding on recent findings with RACS and LOFAR data \citep[e.g.,][]{2021Galax...9...99A,2023A&A...672A.163O,2024A&A...691A.185M},
they will be found by EMU in significant numbers \citep[e.g.][Kataria et al. in prep.]{2021PASA...38....8Q,2022MNRAS.512.6104G, Simonte2024}. Having large samples of extreme systems in both hemispheres will be important for exploring any large-scale cosmological implications.

Dying radio galaxies are important for understanding the duty cycle of AGN \citep[][]{2020MNRAS.496.1706S} and, when relaxed, are a good probe of surrounding pressures \citep{2011A&A...526A.148M, 2018MNRAS.480.5286Y, 2019MNRAS.490.5807E}. Radio remnant morphologies are also excellent probes of jet and environment dynamics, but require excellent surface brightness sensitivity due to the rapid fading of remnant lobes \citep[e.g.][Stewart et al. in prep.]{2023PASA...40...14Y,Riseley_2025}. EMU will provide the best constraints on any current jet or hot spot activity in such systems. The bending and distortion of radio galaxy morphologies is an excellent probe of diffuse gas, providing ideal laboratories to explore the physical conditions in the outskirts of clusters, poor groups, and cosmic filaments.
The demonstrated surface brightness sensitivity of EMU is critical for these studies, as is a large sky area due to the rarity of key source populations with short fractional lifetimes. 

\subsection{Galaxy clusters and large scale structure}
Galaxy clusters present opportunities to study large-scale structure evolution, turbulence, cosmic rays, shocks, feedback, and more. They evolve and grow through a variety of processes, including passive accretion of gas, consumption of small galaxy groups, and violent merger events. Many clusters host vast and enigmatic diffuse radio continuum sources such as giant and mini radio halos, radio relics from merger shocks, and numerous AGN and remnant radio galaxies interacting with the intracluster environment \citep{2019SSRv..215...16V}. Structures on the largest scales, superclusters, are also a likely source of radio emission from filaments and from various cluster merger signatures \citep{
2022A&A...661A..46V,2023JApA...44...38P}.
EMU has already produced results studying such features, with \citet{2020ApJ...900..127H} and \citet{2021PASA...38....5D} exploring diffuse radio sources in the massive merging clusters SPT-CL~J2023$-$5535 and SPT-CL~J2032$-$5627, respectively, \citet{2021A&A...650A.153D} investigating the possibility of a halo in SPT-CL~J2106$-$5844 to extend the detection of radio emission in clusters to $z\gtrsim0.8$, and \citet{loi23} discovering an unusual extended radio arc in Abell 3718. \citet{2023A&A...677A.188B} found a correlation between the radio luminosity of cluster central AGN and the X-ray luminosity of the clusters in the first EMU Pilot Field. Early EMU/\textit{eROSITA} results on the Abell 3391/95 galaxy cluster system \citep{2021A&A...647A...2R, 2021A&A...647A...3B} have helped constrain physical processes in the merger. Detailed analyses of ``clumps'' in the filaments discovered in the Abell 3391/95 system show the influence of the bright wide angle tailed central galaxy \citep{2022A&A...661A..46V}. The complex interplay of radio sources in the spectacular merger systems Abell 3266 and Abell 3627 has been constrained using EMU data \citep{2022MNRAS.515.1871R, 2024MNRAS.533..608K}. \citet{Macgregor2024} have mapped the cluster emission in Abell S1136 using EMU Early Science data, finding that the diffuse emission breaks up into filaments when seen with the sensitivity and resolution of ASKAP. Diffuse radio sources in clusters are known to have very steep spectra, with spectral indices\footnote{We adopt the convention that spectral index, $\alpha$, is related to flux density, $S$, and frequency, $\nu$, through $S\propto\nu^{\alpha}$.} ranging from $\alpha \sim -1$ to $-3$ depending on the type of source (halos, relics, radio remnants), and it is anticipated that a large number of radio halos and relics will be detected by EMU \citep[e.g.][]{cassano12, nuza17, nishiwaki22,Duchesne2024}

Filaments of the cosmic web exist on scales larger than clusters, and are now being mapped using weak lensing \citep{2024eas..conf.1223H}. Galaxy populations inside and outside of filaments have been compared using stacking analyses \citep[e.g.,][]{2017MNRAS.466.4692K},
but detecting radio emission from the filaments themselves is an emerging field. With the new generation of telescopes, faint bridges of diffuse emission are now becoming detectable in pairs of pre-merging galaxy clusters \citep[e.g.][]{2019Sci...364..981G, botteon20a1758, dejong22, 2023A&A...679A.107B,2024A&A...682A.105P}.
EMU Early Science data \citep{2022A&A...660A..81V} show that detecting emission extending beyond the central regions of clusters with ASKAP is possible. EMU will enable sensitive studies of the synchrotron cosmic web. 

While other instruments, such as the MeerKAT radio telescope \citep{2016mks..confE...1J}, will have comparable sensitivity to diffuse cluster phenomena, they will focus only on targeted regions, sometimes including complementary ASKAP data \citep[e.g.,][]{2024MNRAS.531.3357K}. The large area
proposed for EMU will ensure coverage of a vastly greater number of clusters, and in regions unmatched by such targeted surveys.
Only around 150 clusters have so far been found to host diffuse radio sources, such as halos and relics \citep[see, e.g.,][for recent large collections]{botteon22, knowles22, duchesne21eor, Duchesne2024}. EMU allows for an increase in this number to statistically significant samples (thousands) through maximising the sky area covered, along with
its excellent spectral index precision and sensitivity to diffuse emission.

\subsection{Cosmology and fundamental physics}

\begin{figure*}[hbt!]
\centering
\includegraphics[width=0.85\paperwidth]{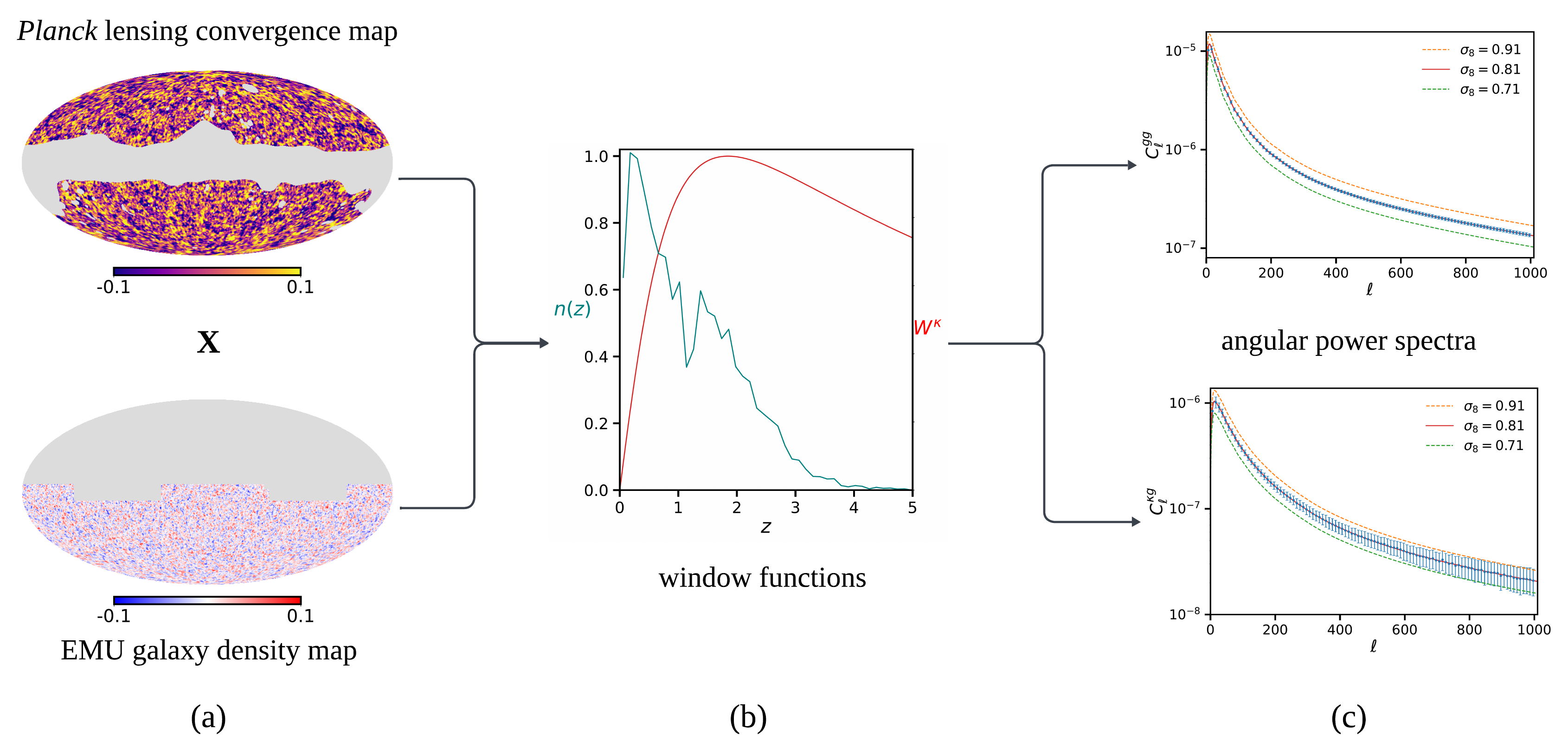}
\caption{A schematic demonstrating how the EMU galaxy density map (a) can be used to measure the auto angular power spectra, and also cross-correlated with other large-scale structure maps (in this case the Planck CMB lensing convergence map, denoted $\textbf{X}$). Combining this with information about the redshift distribution $n(z)$ we can compute the window function (b), and compare the measured auto- and cross-power spectra with their theoretical predictions to constrain the cosmological parameters (in this case the amplitude of the density perturbations $\sigma_8$.}
\label{fig:emu_cosmology_density}
\end{figure*}

EMU has the capability to provide important tests of the fundamental physics of the Universe, including questions such as the nature of dark matter \citep{2021JCAP...11..046R}, the nature of the mysterious force accelerating the expansion of the Universe (dark energy), and the mechanism that generates the initial conditions, both of which are currently unknown. Dark energy is an established part of the cosmological model, but the evidence at low-redshift comes mainly from standard candles and standard rulers. Radio galaxies from EMU can be used to trace the distribution and the evolution of the gravitational potential at higher redshifts through their clustering statistics. By cross-correlating the EMU galaxy distribution with the cosmic microwave background (CMB) the late-time integrated Sachs-Wolfe (ISW) effect \citep{1996PhRvL..76..575C,2012MNRAS.424..801R,2018PhRvD..97f3506S,2022MNRAS.517.3785B} can be measured. Cross-correlating with galaxy surveys at lower redshifts, including samples drawn from the \textit{Wide-field infrared survey explorer} \citep[\textit{WISE},][]{2010AJ....140.1868W}, VISTA hemisphere survey \citep[VHS,][]{2013Msngr.154...35M}, Dark Energy Survey \citep[DES,][]{2016MNRAS.460.1270D}, Kilo-Degree Survey \citep[KiDS,][]{2015A&A...582A..62D}, and the Dark Energy Spectroscopic Instrument \citep[DESI,][]{2016arXiv161100036D}, further allows cosmic magnification \citep{2005ApJ...633..589S} to be detected and used. These data will thus enable a highly sensitive test of General Relativity at large scales. Either of these cross-correlation approaches also allows for the characterisation of the redshift distributions and bias of the radio source population \citep{2021MNRAS.502..876A}. The cross-correlation of the first EMU Pilot galaxy sample with CMB lensing is an illustration of what the full EMU survey will enable for such characterisation studies, and for the measurement of the growth of structures \citep{2024arXiv241105913T}, complementary to weak lensing surveys. EMU data can also be used to expand on an analysis of the cosmic dipole \citep{2024MNRAS.531.4545O} that suggests a tension between the dipoles inferred from the radio source distribution of NVSS and RACS, and the kinematic dipole of the CMB.

The EMU sample will also provide key information to test the inflationary theory by determining the large-scale Gaussianity of the initial distribution of structures \citep{2017PDU....15...35R,2019JCAP...02..030B}. These clustering statistic approaches are complementary to other established cosmological probes (e.g. CMB, type-Ia supernova, and baryon acoustic oscillations and galaxy clusters). However, they require a large contiguous area and high source density for uniform sampling. The limitations of future continuum clustering surveys can be related to the low density of AGNs or intrinsic confusion noise \citep{2021MNRAS.506.4121A}. The cosmic radio dipole, caused by our peculiar motion with respect to the rest frame where EMU galaxies are statistically isotropic, also contributes strongly to the large-scale clustering signal and can mimic that of a non-Gaussian initial distribution of structures \citep{2016A&A...591A.135C}. Previous measurements of the cosmic radio dipole direction \citep{2002Natur.416..150B,2011ApJ...742L..23S,2012MNRAS.427.1994G,2016JCAP...03..062T,2021A&A...653A...9S} are in agreement with the expectation from the CMB \citep{2020A&A...641A...1P}, but their amplitudes are in tension. However, the dipole amplitude from recent MeerKAT absorption line survey data agrees with the CMB after including sub-mJy sources \citep{2024arXiv240816619W}. The sensitivity of EMU will allow testing of this result using a $\sim$5 times larger survey area.

The density of sources in EMU, a factor of 10 or more greater compared to NVSS and RACS, will be significant in addressing other unresolved cosmological questions, with one such being the origin of the CMB cold spot region \citep[$\alpha=03^h~15^m~05^s$, $\delta= -19^{\circ}~35^{\prime}~02^{\prime\prime}$;][]{2007ApJ...671...40R,2010MNRAS.403....2S,Faisal_ur_Rahman_2020}. There is a statistically significant underdensity in radio source numbers and surface brightness in the region of the cold spot, shown by \citet{2007ApJ...671...40R} using NVSS data. They argue that this implies it is cosmologically local, resulting from a localised manifestation of the late-time ISW effect. With increased source density, and redshift estimates for many EMU sources, this result can be explored in more detail to identify the extent, and potentially the redshift localisation, of any underdensity, and the likelihood of association with the origin of the cold spot.

\subsection{The Galaxy and Magellanic Clouds} 
 \label{sec:gal_mag_sci}
EMU will create the most sensitive wide-field atlas of Galactic continuum emission in the southern hemisphere, along with some of the most sensitive maps of the Magellanic Clouds, allowing the study of the formation and evolution of stars in exquisite detail. EMU's observations of the Magellanic Clouds will complement existing MWA \citep{2018MNRAS.480.2743F}, Australia Telescope Compact Array (ATCA) higher-frequency \citep{2002MNRAS.335.1085F,2011SerAJ.183...95C} and similar-sensitivity MeerKAT observations \citep{2024MNRAS.529.2443C,2024MNRAS.531.2835C,2025A&A...693L..15S}, extending measurements of spectral indices for all detected sources. 

EMU will observe the full range of Galactic latitudes, extending well beyond the few degrees on either side of the plane to which many Galactic surveys are limited. 
EMU's sensitivity, especially to extended emission, is needed to increase samples of supernova remnants \citep{2022MNRAS.512..265F,2023MNRAS.518.2574B,2024MNRAS.534.2918S,2024A&A...692A.237Z}, planetary nebulae \citep{2024Ap&SS.369...85A}, and the newly-detected association of low surface brightness H\,{\sc ii} emission with reflection nebulae such as Lagotis \citep{2025PASA...42...32B}, which together support the study of the end stages of stellar evolution.
EMU will reveal all the stages in the evolution of a compact H\,{\sc ii} region (hypercompact, ultracompact and compact).

Early science and pilot data have shown EMU's potential for these studies \citep[e.g.,][]{2019MNRAS.490.1202J,2021MNRAS.506.3540P,2021MNRAS.507.2885F,2021MNRAS.506.2232U,2023MNRAS.524.1396B}. One highlight is the unexpected detection of asymptotic giant branch (AGB) stars in EMU's wide-band images through their OH maser emission, despite these objects lacking continuum emission \citep{2022MNRAS.512L..21I}. The large fractional bandwidth and resolution of EMU, coupled with infrared surveys of the Galactic Plane (VISTA VVV, \textit{Spitzer} GLIMPSE and MIPSGAL, \textit{Herschel} Hi-GAL) allow us to distinguish thermal radio emitters (H\,{\sc ii} regions, planetary nebulae) from non-thermal (supernova remnants, pulsar wind nebulae, pulsars, active stars). High-energy observations (both gamma rays from High Energy Stereoscopic System (HESS) and the \textit{Fermi} Gamma-ray Space Telescope, and X-rays from \textit{Chandra} X-ray Observatory, \textit{XMM-Newton}, the Neil Gehrels \textit{Swift} Observatory, and the \textit{eROSITA} all-sky survey) also provide strong synergies. ASKAP short baselines recover spatial scales up to $\approx 45$\,arcmin at $\approx 1$\,GHz, a unique capability among interferometers, making EMU ideal to study large and complex Galactic structures \citep[e.g.,][]{2021MNRAS.506.2232U,2024PASA...41..112F}.

The Galactic science goals of EMU include: (1)~a complete census of the early stages of massive SF in the southern Galactic Plane; (2)~detection and characterisation of a significant number of missing supernova remnants up to the edge of the Galactic disk \citep[$\sim\!300$ known, up to $2000$ expected, e.g.,][]{2022ApJ...940...63R,2023MNRAS.524.1396B,2023AJ....166..149F,2024RNAAS...8..107L,2024RNAAS...8..158S,2024MNRAS.534.2918S,2024arXiv241103367G,2024A&A...690A.203B,2025ApJ...980..162J}; (3)~understanding the complex structures of giant H\,{\sc ii} regions and the inter-relationship of dust, ionised gas and triggered SF \citep[][Bradley et al., in prep.]{2014AJ....147..162D,2017ApJ...843...61S}; (4)~the variety of luminous blue variables \citep{Bordiu2024}; (5)~serendipitous discoveries, such as the radio flares from ultra-cool dwarfs found by \citet{2001Natur.410..338B}: these stars are distributed isotropically in the sky, and EMU is the only wide-area southern survey planned at this frequency and sensitivity; (6)~a detailed characterisation of planetary nebulae \citep[spectral energy distributions, distances, ionised mass,][]{2021MNRAS.503.2887B}. Detection of other classes of radio stars, adding to the growing numbers recently identified \citep{2023PASA...40...36D,2024PASA...41...84D} will also be enhanced through the sensitivity of EMU.

\subsection{Pulsars, Variables, and Transients}
While pulsars are primarily detected and observed with high temporal resolution in order to resolve their pulses, the phase-averaged emissions of pulsars can also be detected in radio continuum surveys. Deep all-sky continuum surveys like EMU enable the study of the spectral, polarisation and scintillation properties of a large sample of radio pulsars \citep{2016MNRAS.461..908B,2017PASA...34...20M,2023ApJ...956...28A,2024arXiv240713282S}. EMU will also enable us to carry out targeted searches for radio pulsars over the whole visible sky while avoiding the need for expensive pixel-by-pixel searches with high temporal resolution, strongly complementing other pulsar surveys in the southern hemisphere with MWA \citep{2023PASA...40...21B,2023PASA...40...20B}, MeerKAT \citep{2023MNRAS.524.1291P} and the phased array feed on Murriyang/Parkes \citep{2016ceaa.conf..909C,2017PASA...34...26D}. Continuum surveys are equally sensitive to all pulsars, not affected by the dispersion measure smearing, scattering or orbital modulation of spin periods. This allows for the discovery of extreme pulsars such as sub-millisecond pulsars, pulsar-black hole systems and potentially also pulsars in the Galactic Centre \citep[e.g.,][]{2024ApJ...967L..16L}. The capability of finding new pulsars with ASKAP has been demonstrated by the discovery of pulsars originally identified as highly polarised radio sources \citep{2019ApJ...884...96K,2024ApJ...961..175W} and pulsars associated with SNRs \citep[e.g.,][]{2025MNRAS.537.2868A} and pulsar wind nebulae \citep{2024PASA...41...32L}. The deeper observations of EMU have the potential to reveal a large population of these sources, especially millisecond pulsars at intermediate and high Galactic latitudes. To distinguish pulsars from other point sources \citet{2016MNRAS.462.3115D} developed a formalism for computing variance images from standard interferometric radio images and demonstrated its feasibility. \citet{2017MNRAS.472.1458D} showed that, with the variance imaging technique alone, EMU should discover $\sim\!40$ new millisecond pulsars and $\sim\!30$ new normal pulsars. 
Variance imaging with EMU will be more sensitive than current pulsar surveys at high Galactic latitudes.

Searches for radio transients and variable sources have previously been limited by small fields of view and poor sensitivity \citep[e.g.,][]{laz+10,oben+14}. The ASKAP telescope is conducting the Variables and Slow Transients \citep[VAST,][]{2013PASA...30....6M,2021PASA...38...54M} survey specifically to improve this parameter space. EMU data provide a complementary resource to VAST in identifying transients on shorter timescales. 
With its large sky coverage, EMU enables a large-scale radio transient and variability search on shorter timescales ($<1$\,h), largely unexplored before \citep[e.g.,][]{tao+23,wang+24}. Studies on 15 minute timescales using the first EMU Pilot Survey have identified 11 such sources and the full EMU survey may identify up to $\sim$1000 variable sources \citep{wang+23}. More recently, several ultralong period (ULP) sources, with periods of several minutes, with repeating bursts of coherent radio emission have been reported \citep[e.g.,][]{caleb+22,walker+22,walker+23,dong+24}. ASKAP has demonstrated its capability to discover such sources \citep[][]{caleb+24,dobie+24} and the full EMU survey is expected to discover many more. 

EMU has also been used to search for a radio-continuum counterpart of the recent ultra-high-energy (UHE) neutrino event, KM3-230213A \citep{2025Natur.638..376K}. Among more than 1000 radio sources within the 68\% confidence region of the UHE neutrino event, three distinctive radio sources, a nearby spiral galaxy (UGCA~127), a radio AGN, and a compact variable radio source (blazar), were found to be possible origins for KM3--230213A \citep{2025arXiv250309108F}.

\subsection{Resolved Galaxies in the Local Universe}
\begin{figure*}[ht]
\begin{center}
\includegraphics[width=0.99\textwidth]{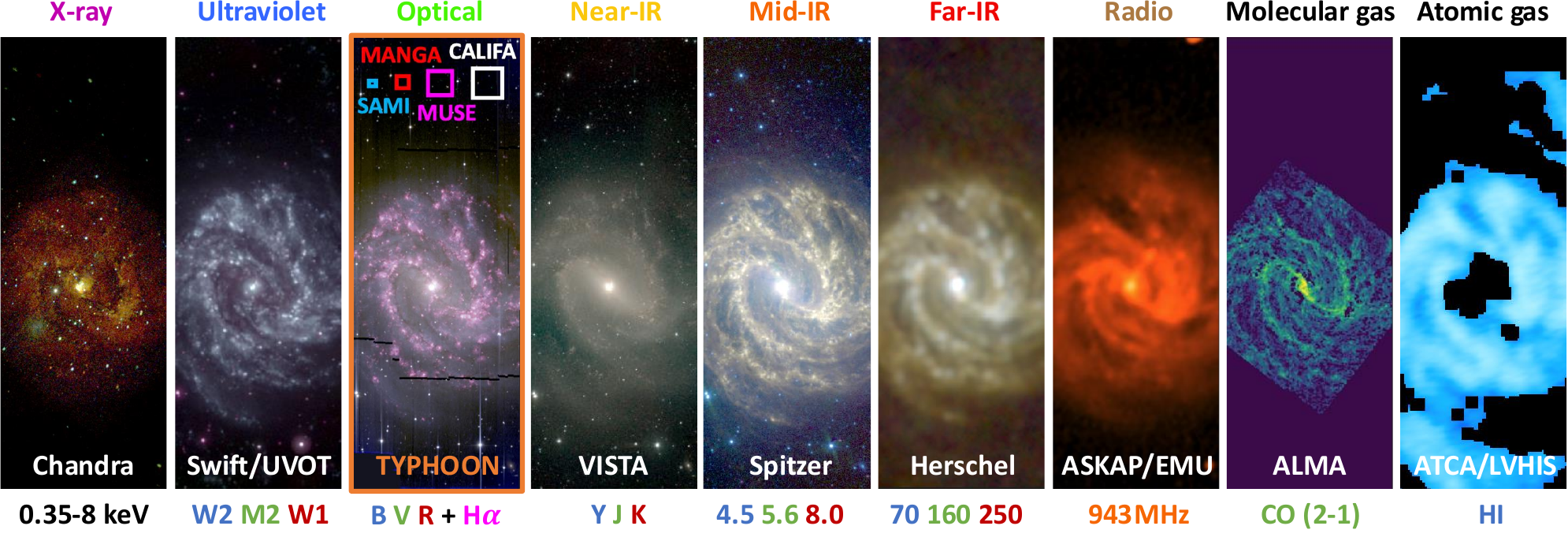}
\end{center}
\vspace{-6mm}
\caption{Demonstration of the exquisite ancillary data available for the nearby spiral galaxy M\,83. The filters used to create each colour image are indicated on the bottom. 
The size of the panels were matched to the TYPHOON survey field-of-view (orange box; third panel), which is $\sim$100--1000$\times$ larger relative to other optical IFS instruments (smaller boxes). We note a MUSE mosaic of 26 tiles is also available for this galaxy \citep{dellaBruna22}. The ASKAP/EMU data (seventh panel) are from a single observing block ($5\,\text{h}$, half of the total final integration time). Similar nearby galaxies with extensive multiwavelength photometry and spectroscopy offer a unique opportunity for spatially resolved comparisons between all baryonic components within galaxies (stars, ionised gas, molecular gas, atomic gas, and dust). The ancillary data are retrieved from the following: \textit{Chandra} -- Harvard/SAO; \textit{Swift}/\textit{UVOT} -- NASA HEASARC service; TYPHOON -- priv. comm.; VISTA -- ESO science archive; \textit{Spitzer} and \textit{Herschel} -- NASA Extragalactic Database; ALMA -- PHANGS-ALMA \citep{leroy21}; ATCA/LVHIS \citep{Koribalski2018}. Higher resolution H\,{\sc i} data from ASKAP/WALLABY for M\,83 will be available in the future. \label{fig:typhoon_demo}}
\end{figure*}

The sensitivity and resolution of the EMU survey enable detailed, spatially-resolved studies of several thousand nearby galaxies ($D\lesssim50$\,Mpc).
EMU's resolution of $\sim15''$, is well matched to the resolution of the \textit{WISE} W4 band (22\,$\mu$m), providing $\gtrsim$\,5 synthesised beams (resolution elements) across galaxies with $M_B<-18$\,mag ($\log(M_\star/M_\odot)\gtrsim9$) at $D\lesssim 50$\,Mpc \citep{leroy19}.
The EMU data are well-suited for comparison with existing atlases of thousands of nearby galaxies from all-sky imaging surveys, such as GALEX \citep[UV; e.g.,][]{gildePaz07}, 2MASS \citep[near-IR; e.g.,][]{jarrett03}, and \textit{WISE} \citep[mid-IR; e.g.,][]{jarrett19, leroy19}. A smaller subset of galaxies ($\sim50$) could also be compared to optical integral field spectroscopic surveys, such as PHANGS-MUSE \citep{emsellem22}, TYPHOON (Seibert et al. in prep; see Figure~\ref{fig:typhoon_demo}), and SDSS-V Local Volume Mapper \citep[LVM,][]{drory24}.

In star-forming galaxies, most of the synchrotron emission at $\sim\,1\,$GHz comes from cosmic-ray electrons accelerated by SNRs from high mass ($M_\star>8\,M_\odot$) stars, with lifetimes of $\tau\lesssim50\,$Myr. The typical lifetimes of cosmic-ray electrons are $\tau\lesssim100$\,Myr \citep{condon&ransom16}, and as a result, radio continuum emission from normal galaxies traces recent SF on similar timescales.
Understanding the exact SF timescale traced by synchrotron emission has been difficult to pin down through global measurements of galaxies \citep[e.g.,][]{cook24}.
By linking resolved EMU data with independent tracers of SF at other wavelengths, such as UV, optical (emission lines), and infrared \citep{kennicutt&evans12}, it is possible to explore this link in fine detail. Closely related to the link with SF, star-forming galaxies also show a well-known correlation between their far-IR and radio emission \citep[e.g.,][]{condon91, yun01}. The physical origin of this correlation is poorly understood and will also be studied in greater detail through such spatially-resolved comparisons of a subset of galaxies with available far-IR data \citep[e.g.,][]{kennicutt11}.

Additionally, EMU has access to continuum observations with ASKAP Band 2 (at 1.4\,GHz), that are commensal with the WALLABY
Survey (see \S\,\ref{sec:wallaby}). Early results from WALLABY's continuum observations of the Eridanus supergroup demonstrated the potential for the better resolution ($6.1'' \times 7.9''$) continuum observations of Band 2 to be useful for probing the internal physical mechanisms that are occurring in star-forming regions in combination with ancillary observations at shorter wavelengths \citep{2023PASA...40...12G}. In the nearby galaxy IC~1952, \citet{2023PASA...40...12G} found that the resolved infrared-radio correlation could be used to distinguish between a background AGN (near the plane of the galaxy) and star-forming regions within the galaxy. Disentangling the AGN and SF properties within galaxies is critical in understanding not only the multiwavelength tracers of SF but also any feedback links between the two.

\subsection{Discovering the unexpected}
\label{sec:discovery}
Experience has shown \citep{2017PASA...34....7N} that whenever we observe the sky to a significantly greater sensitivity, or explore a significantly new volume of observational phase space, we make new discoveries. Even the Australia Telescope Large Area Survey (ATLAS) \citep{2006AJ....132.2409N,2008AJ....135.1276M}, which expanded the phase space of sensitive radio surveys by only a factor of a few, discovered two previously unrecognised classes of object (infrared faint radio sources, and radio-loud AGN buried in SF galaxies). With a sensitivity 10-30 times better than other large-scale radio surveys at similar frequencies, EMU is likely to discover new types of object, or new phenomena. While impossible to predict their nature, we might reasonably expect new classes of galaxy or new Galactic populations.

This goal has already been demonstrated in EMU through the successful identification of a new class of radio object, odd radio circles \citep[ORCs,][]{2021Galax...9...83N,2021PASA...38....3N,2021MNRAS.505L..11K,2022PASA...39...51G}. These diffuse circles of radio emission are now known \citep{2022MNRAS.513.1300N}, at least in some cases, to surround galaxies at $0.2 < z < 0.6$. Other patches of diffuse emission which may be related to ORCs have also been found around galaxies \citep{2024MNRAS.532.3682K} or groups of galaxies \citep{2024A&A...685L...2B}.
The cause of these circles is still debated, but may result from a starburst wind termination shock from the central host galaxy \citep{2021PASA...38....3N}, from a spherical shock wave from a cataclysmic event in the central host galaxy \citep{2021PASA...38....3N}, from a shock in a galaxy-galaxy merger event \citep{Dolag2023-sims}, or from re-energised electrons in a relic radio lobe \citep{shabala24}.
The combination of their rarity (about one per 100\,deg$^{2}$) and low surface brightness explains why they have not been discovered before. We expect to discover about 300 ORCs in the full EMU survey, making them an important new class of radio objects.

To expedite the discovery of ORCs and other unusual radio morphologies, we have developed ML algorithms that leverage multi-modal foundation models. These models use text and image embeddings to retrieve similar radio sources from the first year of the EMU full survey in under a second. To make these models accessible, we created the EMU Survey Object Search Engine (EMUSE\footnote{\url{https://askap-emuse.streamlit.app/}}, Gupta et al., submitted), which enables users to search for similar objects using text descriptions or image uploads.
In another work (Gupta et al., submitted) we present the ORCs and other unusual sources identified using a combination of object detection methods from \citet{2024PASA...41....1G} and these foundation models, from data observed during the first year of the full EMU survey.

There are many additional areas in which EMU data will support novel discoveries. These include the detection, measurement, and exploration of binary supermassive black holes; populations of gigahertz peaked spectrum (GPS), compact steep spectrum (CSS), and other young AGN to understand the AGN duty cycle and evolution \citep[e.g.,][]{2017ApJ...836..174C}; spatially resolved nearby galaxies to explore SF and extreme stellar evolutionary states; properties of dark matter \citep{2017ApJ...839...33S}; the stellar initial mass function and whether, and how, it may evolve \citep[e.g.,][]{2018PASA...35...39H}; and more.
The discovery of ORCs has demonstrated the value of exploring new regions of observational parameter space. We expect that the full EMU survey will continue to reveal new and rare classes of objects as we maximise the sky areas probed to this unprecedented sensitivity. A few additional examples of objects for which current models are inadequate are presented in Section \ref{sec:unexpected}.

\subsection{Legacy}
\label{sec:legacy}
EMU was always conceived as a project with immense legacy value, in addition to the direct scientific outcomes planned. As now shown through more than three decades of major sky survey projects across the electromagnetic spectrum, each supports a wealth of science beyond that anticipated, and underpins further projects as new complementary surveys arise.

EMU will be the touchstone $\sim\!1\,$GHz radio continuum atlas for at least a decade after completion, and possibly beyond. While the SKA's survey plans are not yet finalised, its strengths suggest it may prioritise a wedding-cake-tiered continuum survey program with the widest tier limited to smaller sky areas ($\sim\!1000\,$deg$^2$). Consequently, no planned survey would replace EMU for the foreseeable future.
EMU will provide the most uniform deep wide-area radio data at GHz frequencies to complement galaxy and AGN evolution, physical processes and cosmological studies from recent and upcoming surveys and facilities such as \textit{SRG}/\textit{eROSITA} \citep[][]{Merloni2024,Predehl2021}, DES, DESI, GLEAM-X, the LOFAR Two-metre Sky Survey \citep[LoTSS,][]{Shimwell2019}, PanSTARRS, Apertif, 4HS, WAVES, the Rubin Observatory's Legacy Survey of Space and Time \citep[LSST,][]{2019ApJ...873..111I}, \textit{Euclid}, and more. A recent example of this radio, optical, X-ray complementarity using EMU data is the discovery of radio emission from a 15\,Mpc filament in the intergalactic medium, a warm gas bridge, infalling matter clumps, and (re-)accelerated plasma in the Abell 3391/95 galaxy cluster system \citep{Reiprich2021}.

EMU is already demonstrating its capacity to serve as a critical and key data resource in ML algorithm development for radio astronomy. The ML applications for EMU range across source finding, host or multiwavelength cross identification, identification of multi-component or giant radio galaxies, image denoising, redshift estimation, and more. Such algorithms will be applicable to many future datasets, especially those delivered by the SKA. In addition to ML, EMU is demonstrating other new software development, and providing an example of unifying survey pipeline, data hosting, value-added data development and delivery, across multiple organisations and technologies. The legacy of software, tools, and data infrastructure and infrastructure linkages developed for the specific needs of EMU will also inform future large scale astronomical developments, in particular for the SKA.

\section{SURVEY DESIGN}
\label{sec:survey}
To determine the survey sky coverage, the EMU and POSSUM teams explored several options that can satisfy all constraints set by the observatory, as well as by both science teams. These constraints are:
\begin{itemize}
\item A $2\pi$\,sr areal coverage, decided by the telescope time awarded to EMU and POSSUM;
\item A contiguous sky region, as any missing sky coverage due to having gaps or holes would be problematic for cosmology analyses looking for structures on large angular scales;
\item Maximal overlap with equatorial multiwavelength and spectroscopic sky surveys, especially the SDSS Extended Baryon Oscillation Spectroscopic Survey (eBOSS) emission line galaxy (ELG) cosmology programme \citep{2020ApJS..249....3A}, the GAMA survey \citep{2022MNRAS.513..439D}, and the WAVES survey \citep{2019Msngr.175...46D};
\item Overlap with complementary radio surveys of the northern sky, especially LoTSS and the Very Large Array Sky Survey \citep[VLASS,][]{Lacy2020};
\item Coverage of the southern Galactic Plane, as well as the Magellanic System, to enable science at good spatial resolution in our cosmic neighbourhood;
\item Coverage of the extreme southern sky ($\delta < -40^\circ$) that cannot be accessed by telescopes in the northern hemisphere, to enable discovery in these less-explored sky areas.
\end{itemize}

\begin{figure}
    \centering
    \begin{subfigure}[b]{1\linewidth}
    \includegraphics[width=0.95\columnwidth, trim=50 50 50 0,clip]{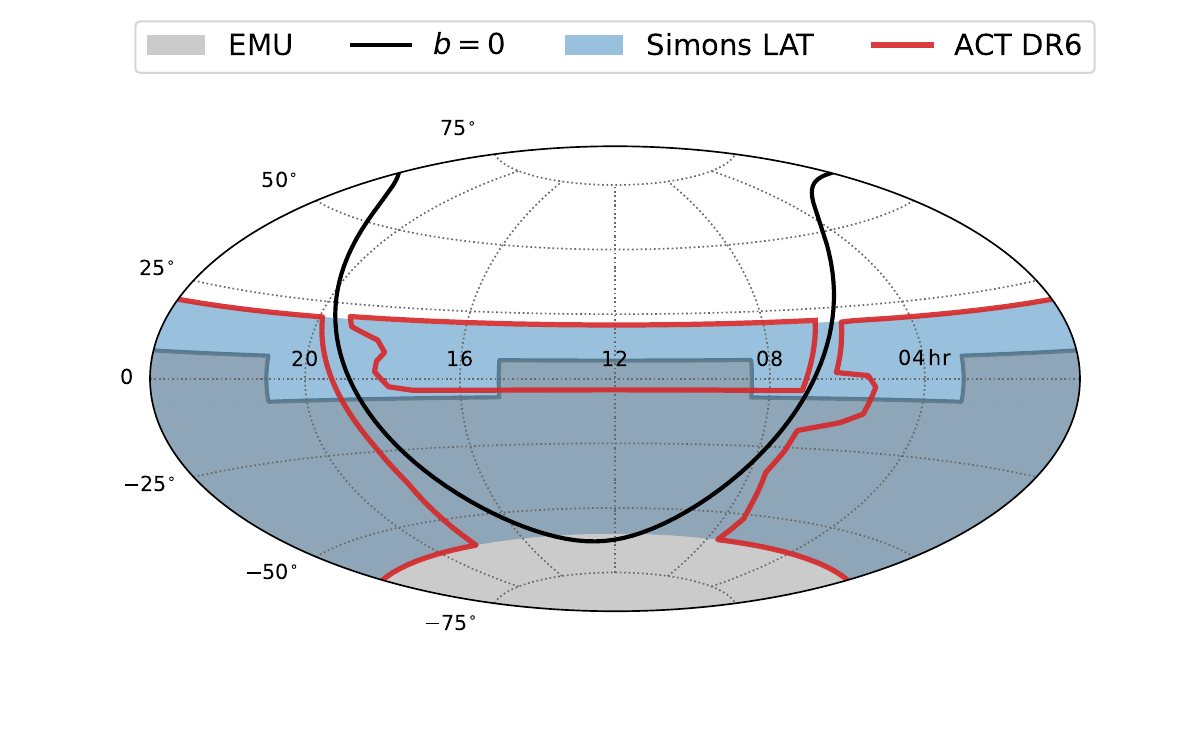}
    \caption{\label{fig:mw:mm}}
    \end{subfigure}\\%
    \vspace{2em}
    \begin{subfigure}[b]{1\linewidth}
    \includegraphics[width=0.95\columnwidth, trim=50 50 50 0,clip]{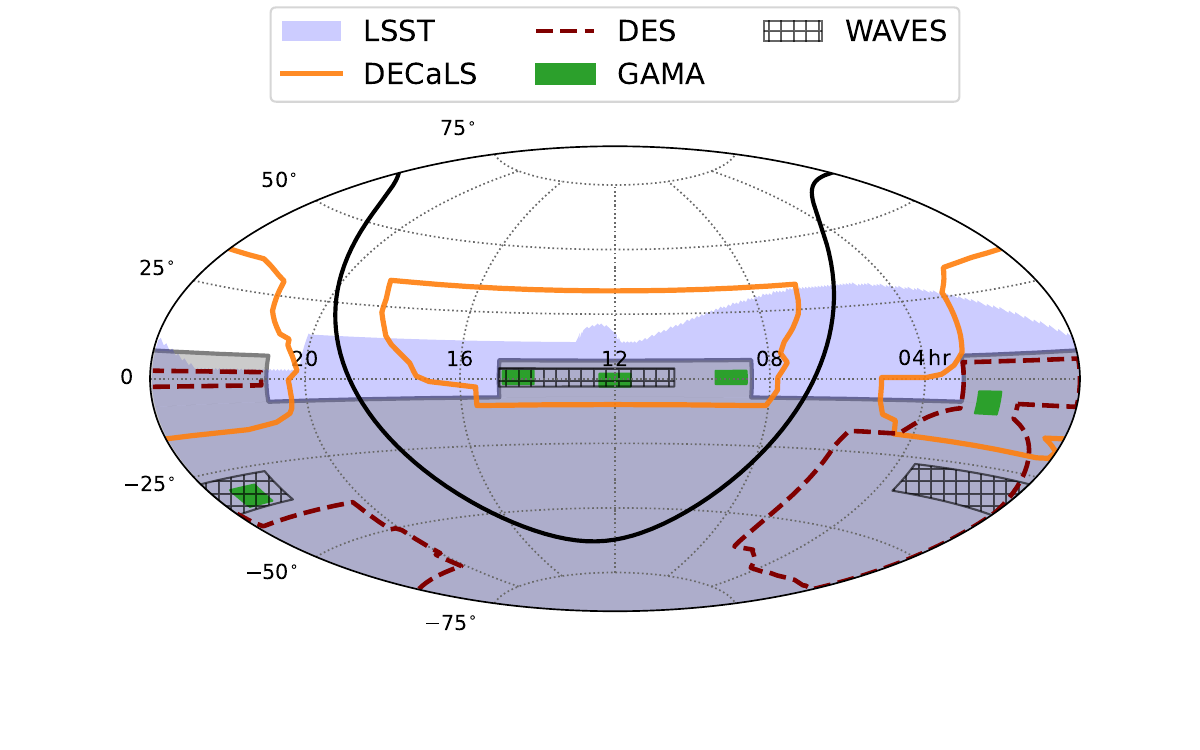}
    \caption{\label{fig:mw:optical}}
    \end{subfigure}\\%
    \caption{\label{fig:mw} 
    Some examples illustrating the overlap in sky coverage of EMU with (a) mm and (b) optical surveys.
    Panel (a) shows the footprint of the Atacama Cosmology Telescope 6th Data Release cluster survey (red outline) and the expected coverage of the forthcoming Simons Observatory Large Aperture Telescope survey \protect\citep[blue shading, ][]{2019JCAP...02..056A}.
    Panel (b) shows the survey footprints for DES \protect \citep[dark red dashed outline,][]{2016MNRAS.460.1270D}, DECaLS \protect \citep[orange solid outline,][]{2019AJ....157..168D}, LSST \protect \citep[blue shaded region,][]{2019ApJ...873..111I}, GAMA \protect \citep[green shaded region,][]{2011MNRAS.413..971D, 2015MNRAS.452.2087L} and WAVES Wide \protect \citep[black hatched region,][]{2019Msngr.175...46D}.
    In both panels the EMU survey footprint is shown by the grey shaded region and the Galactic equator by the black line.}
\end{figure}

Input from the EMU, POSSUM, and WALLABY teams was solicited, leading ultimately to the current EMU footprint shown in Figure~\ref{fig:coverage}. This option met POSSUM's primary goal of ensuring a large contiguous areal coverage with minimal adjustment for specific targets, as well as EMU's goal of contiguous sky coverage encompassing suitable complementary multiwavelength and spectroscopic datasets (Figure~\ref{fig:mw}). It also encompasses the full area of the WALLABY survey.

\begin{figure}[hbt!]
\centering
\includegraphics[width=0.95\textwidth]{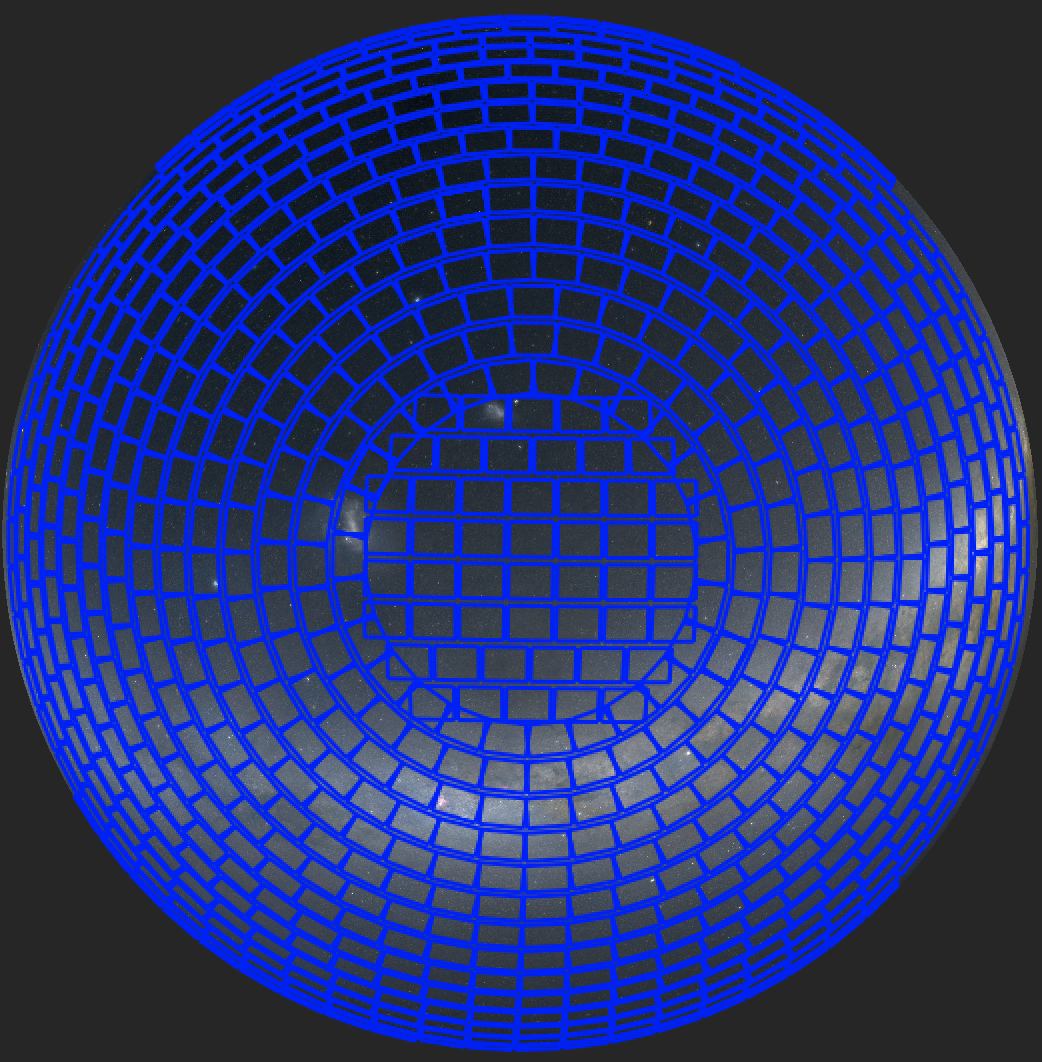}
\caption{The tessellation selected to cover the celestial pole, showing the change in strategy at $\delta<-70$. The blue outlines correspond to each EMU tile.}
\label{fig:pole}
\end{figure}

Early in the development of the EMU survey strategy, it was recognised that optimal tiling of the sky would require different choices for the polar cap compared to the bulk of the rest of the survey. For most declinations, bands of tiles at a fixed declination work well in tiling the sky. This approach, however, is not efficient for regions close to the celestial pole. For declinations $\delta<-70^{\circ}$ a rectilinear grid centred on the pole was chosen (Figure~\ref{fig:pole}). This choice leads to some increased overlap between adjacent tiles at the boundary between the two tiling models, but remains an optimum choice to retain minimal overall duplication, while also facilitating practical telescope operations.
The final EMU sky coverage is produced through a tessellation of 853 ASKAP tile footprints. Each requires a $10\,\text{h}$ integration to achieve EMU's sensitivity goals. Of the 853 total regions, 161 lie at declinations far enough north ($\delta > -10^{\circ}$) that a full $10\,\text{h}$ track is not feasible in a single observation. These tiles are split into two separate ($5\,\text{h}$) integrations. Each EMU tile is given a name in the format of EMU\_hhmm$\pm$dd. For those tiles split into two $5\,\text{h}$ integrations, the two unique names given are EMU\_hhmm$\pm$ddA and EMU\_hhmm$\pm$ddB.

\begin{figure}[hbt!]
\centering
\includegraphics[width=0.95\textwidth]{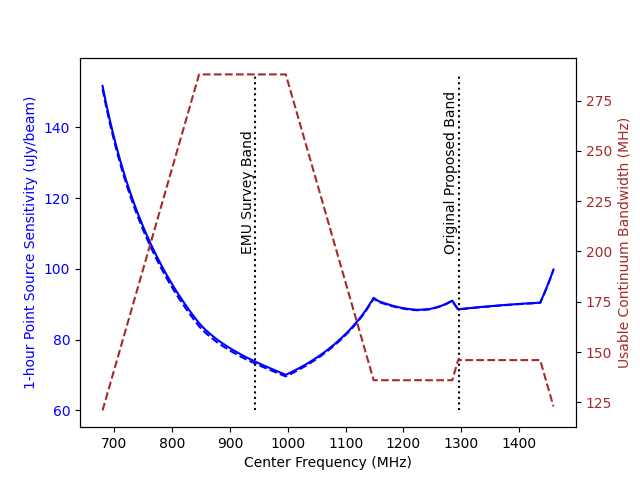}
\caption{ASKAP sensitivity and usable bandwidth as a function of frequency. The centre frequency refers to the middle of the 288\,MHz of correlated bandwidth prior to rejection of unusable channels. The point source sensitivity includes estimates of robust weighting, convolution to a common restoring beam and a mosaicing factor.}
\label{fig:sensitivity}
\end{figure}

In the course of ASKAP Early Science and Pilot science observations, it was established that a central frequency of 943\,MHz would be the optimum for EMU. This choice provides the best point source continuum sensitivity while retaining the maximal bandwidth by avoiding the RFI at higher frequencies (Figure~\ref{fig:sensitivity}).

\section{OBSERVATIONS AND DATA PROCESSING}
\label{sec:obs}
EMU observations are optimised with all 36 ASKAP antennas, but can be scheduled with 32/36 (or 34/36 for more northerly tiles), to meet the required $uv$ coverage and sensitivity goals. EMU uses a central frequency of 943\,MHz with a bandwidth of 288\,MHz. Each tile observed by ASKAP is allocated as a ``scheduling block'' (SB) with an associated unique ID number.
Autonomous scheduling of observations is carried out by an observatory-managed tool called SAURON \citep[][Moss et al. in prep]{2022scio.confE..20M}, which takes into account a variety of survey, environmental and system constraints. These include allocation of time to each of the ASKAP surveys in appropriate proportion, consideration of tile availability and observing constraints (such as distance to the Sun), observing program prioritisation for different tiles or tile groups, and more. As a consequence, while still deterministic, the scheduling is dynamic and complex. The motivation for this approach is to maximise the survey efficiency of ASKAP while balancing many competing factors and operating in a remote and extreme environment.
In addition to these constraints, EMU's observations aim to prioritise tiles adjacent to previously completed tiles, in order to build up locally contiguous sky coverage in regions in order to facilitate the triggering of the The EMU Radio and Value-added Catalogue (EMUCAT) pipeline (See \S,\ref{sec:emucat} below).

The data are processed using the ASKAPsoft pipeline, including multi-frequency synthesis imaging, multi-scale clean and self-calibration.
The process is largely the same as that detailed by \citet{2021PASA...38...46N} for the first EMU Pilot survey. In particular, the clean scales and weighting scheme are the same. Some improvements include the fact that the cleaning now goes down to a $5\sigma$ level, to take into account varying image quality, rather than using the absolute threshold adopted earlier, as that could potentially result in over- or under-cleaning. In addition, the
$w$-term handling has been improved, to allow almost twice the number of $w$-planes. Phase self-calibration is now done initially using a sky model derived from RACS \citep{2020PASA...37...48M, Hale2021} to fix phases over the full integration so that deconvolution has a good starting point. No amplitude self-calibration is performed. The full sequence starts with self-calibration against RACS, an initial deconvolution step, then another self-calibration against the deconvolution model, followed by a final stage of deconvolution.
For details of the array configuration, justification of the choice of baseline weighting, and other technical considerations of ASKAPsoft, see \citet{2019ascl.soft12003G} and \citet{2020ASPC..522..469W}.
After validation (see \S\,\ref{sec:validation} below) EMU data become available on CASDA. EMU is identified as project AS201, WALLABY as project AS202, and POSSUM as project AS203 on CASDA.
The publicly available EMU data include three versions of each validated tile mosaic for each of Stokes I and V. The Stokes Q and U datasets are available as POSSUM data products. The three EMU image versions are produced from the same calibrated dataset, but with different choices of restoring beam for the imaging. The recommended data product includes ``conv'' as a suffix in the filename, and has been restored with a common $15''$ resolution for each beam in the mosaic of a tile. This is consistent for all tiles across the full survey area, and these data are the source for self-consistent flux density measurements across the full survey area. The other two options have the suffixes ``raw'' and ``highres''. The ``raw'' image comes from using the native restoring beam, dependent on non-flagged $uv$ sampling for each beam in the mosaic. This provides a better resolution than the ``conv'' images, typically around $11-13''$ but will have varying resolution across the mosaic, and will also differ from tile to tile. Flux densities derived from these images are not necessarily self-consistent between tiles, or even within the same tile mosaic. The ``highres'' image, made with uniform weighting, is restored with a smaller synthesised beam again, typically $7-9''$. This is consistent across each beam within a tile mosaic, but will differ from tile to tile, and these images have somewhat poorer sensitivity and imaging fidelity compared to the others. Their value is in highlighting finer-scale structure for (typically brighter) objects of interest, while coming at the expense of losing sensitivity to extended emission. Flux densities in these images, again, will not necessarily be self-consistent.

The reason for choosing the common resolution of $15''$ in the ``conv'' images is a consequence in part of the weighting scheme chosen to maximise sensitivity to faint sources, but is primarily driven by the resolution limitations of tiles close to the equator, where ASKAP's $uv$ coverage is somewhat poorer than in tiles further south. This is necessary to ensure a common resolution over the full sky area covered by EMU.

Source catalogues for these images are generated using the {\em Selavy\/} source finder \citep{Whiting2012}. 
{\em Selavy\/} estimates local noise statistics within a sliding box to set its detection threshold, accounting for noise variations across an image. It constructs two catalogues for each tile, an ``island'' and a ``component'' catalogue. The island catalogue is formed by identifying contiguous regions of emission, or islands, above the local background threshold. The component catalogue is an attempt to resolve blended or complex sources, using a sub-thresholding technique within each island. A series of descending flux thresholds are used to identify distinct peaks, which serve as initial guesses for Gaussian component fitting. The resulting component catalogue consists of the fitted Gaussians within each island, constrained by criteria such as minimum separation, positivity, beam-size limits, and consistency with the island’s total flux. A comparison of the performance of {\em Selavy\/} with other point source finders is presented by \citet{2023PASA...40...27B}.

In addition to the EMU images and source catalogues a selection of associated data are available, including the visibilities and a series of ancillary datasets. Ancillary data include weight images used in the mosaicing, clean model images generated during deconvolution, and a series of images generated during source finding, such as background noise images, images containing detected components, and residual images derived from the component image.

\subsection{Data Validation}
\label{sec:validation}
The ASKAP operations team delivers fully processed data cubes from the telescope to the science teams, making use of the Pawsey Supercomputing Centre\footnote{\url{https://pawsey.org.au}}. In order to ensure consistent quality control, the science teams are responsible for validating the data. Given the scale and logistics involved in operating a major survey facility like ASKAP, and the limited size of the operations team, it is impractical to manually reprocess datasets if they have errors. Instead, it is easier and more streamlined simply to reobserve such a tile. In order to decide whether a tile has or has not met the required quality level, the science team is responsible for a process of data validation, which either accepts the tile as meeting the science requirements, or rejects it. The data from rejected tiles are discarded, and the tiles are placed back in the ASKAP observing queue.

Common problems that may lead to errors in the processed data, and which are typically rectified by this reobservation strategy, include excess solar interference, system issues, or erroneous calibrations (bandpass, holography, or, for some observations early in the process, other key datasets that either had errors or that were not observed close enough in time to the dataset in question). These issues are typically flagged in a processed EMU mosaic through excessively large positional or flux density offsets between measured sources and a reference dataset (such as RACS), unusual source count properties, or excessive imaging artifacts (radial spikes, especially around relatively faint sources, or large scale ripples). To facilitate this validation process, each ASKAP science team has established its own validation workflow, which produces a series of diagnostic figures, statistics, and traffic-light metrics to aid the team in rapidly identifying problematic data cubes; EMU makes use of the ASKAP continuum validation workflow\footnote{\url{https://github.com/Jordatious/ASKAP-continuum-validation}}.

ASKAP delivers three to four newly observed and processed tiles for EMU per week, on average, which each require validation.
The validation team includes more than 30 volunteers from among the EMU collaboration, and uses a roster system to share the load of conducting the manual validation and inspection process. In order that the process is as streamlined as possible, a validation workflow was established, with examples and clear step-by-step instructions. This workflow was explained and demonstrated through several training sessions. The validation workflow includes independent recording of key validation metrics, to facilitate subsequent analysis of the EMU tile properties. To ensure a level of consistency in this quality control, the validation team are instructed to escalate potentially problematic tiles to the validation lead and management team for final review before any tile is rejected in the validation process.

The number of rejected tiles has decreased with time as ASKAP operations have become more streamlined, although a significant spike in these numbers occurred during a recent peak in solar activity and associated solar storm events in mid-2024. Overall, about 20\% of EMU observations to date have resulted in rejected tiles through this validation process. Overheads associated with operational improvement over time have been factored into the timing of ASKAP's 5-year observing plan, which assumes an observational efficiency of 70\%.

\subsection{Known image processing limitations}
The ASKAPsoft pipeline needs to accommodate requirements for each of the ASKAP surveys, and in consequence there are inevitably some limitations in the processing that affect the EMU imaging. We outline the main issues and their impact here.

One key issue is a limited implementation in the handling of the $w$-term\footnote{This refers to techniques used to account for the effects of the third spatial coordinate, $w$, in the $uvw$-coordinate system, which represents the baseline geometry in Fourier space.}. As a result, sources near to or outside the edge of the primary beam's FWHM frequently show radial spike or grating-ring type artifacts. Bright sources outside of the primary beam image result in residual grating-ring artifacts and calibration errors (since such sources are not included in the deconvolution during phase self-calibration). There are a number of approaches under development to improve the way this is handled within ASKAPsoft, although these are not yet implemented for existing EMU data products at the time of writing. This includes the introduction of source peeling, and improvements to $w$-term handling, supported by better memory utilisation. One recent straightforward approach, shortly to be added to the official ASKAPsoft pipeline, draws on a sky model for the location of known bright sources. Such bright out-of-field sources are imaged, modelled, and subtracted from the $uv$ data before the main imaging steps to reduce the impact of their associated artifacts. This will serve to provide some improvement in image quality for newly observed EMU tiles.

Apart from initial phase calibration as a result of the bandpass observation, no subsequent phase referencing is performed. This is partially compensated for with phase self-calibration, but relies on a good deconvolution model. Any errors in the deconvolution model will result in calibration errors and subsequent imaging artifacts. Significant astrometric errors can also result when the phases are greatly disturbed over the course of the observation (these are not necessarily consistent from beam to beam and so can result in a slight smearing of sources when overlapping beams are mosaicked). This astrometric error is independent of the signal-to-noise (S/N) ratio of the source (further details are presented in \S\,\ref{ssec:astrometry} below). Investigations of RACS observations have shown that the bulk of the astrometric error is set during the bandpass observation, as this procedure currently takes over two hours to complete (with an approximately 3 minute scan performed on the calibrator source PKS1934$-$638 in each beam). One option to improve the initial bandpass is to either fix the phases against a reference field (and so align the beams) or to perform a shorter field-based bandpass observation and so determine phase solutions for all beams and antennas simultaneously. A longer-term solution being considered by the observatory is to phase-reference the target field against the RACS sky model. This allows time-dependent phases to be corrected as well.

Solar interference has become problematic with a more active Sun. Solar flares are particularly difficult to deal with as they are exceptionally bright, often highly variable and, owing to their compact nature, affect all baselines. A secondary effect is that Solar activity can shift the true beam positions during the beam-forming process since the Sun is used to determine beam weights. In general though, these are expected to be corrected by the holography which is performed for each beam-forming session.

Telescope pointing errors can result in direction-dependent errors, and increased artifacts at the edge of the beam. These are not corrected for by the third telescope axis (which maintains the orientation of the beam with respect to the sky). They not only affect the target source but also the effectiveness of the holography. Further investigation by the observatory team is underway to determine the magnitude and nature of these pointing errors.

Changes in antenna bandpass can detrimentally affect imaging quality. This can occur where an antenna exhibits a partial failure that is not immediately detected by the system and causes amplitude or phase changes in the bandpass. Such an occurrence is rare and is more easily dealt with during the validation process. A lower level issue that affects amplitude gain has been observed to vary diurnally, possibly being temperature related. This is also currently uncorrected for (in simple arrays this would typically be corrected via phase referencing) and could cause low-level artifacts in imaging.

Overall, while there is clearly still room for improvement in the telescope performance and the delivered image quality, these issues do not significantly compromise the integrity of the EMU data provided. Ongoing developments are expected to provide incremental improvements to the EMU data quality of the remainder of the survey.

The EMU team has a goal by the end of the survey to reprocess early tiles to bring them into line with the final image quality to homogenise the sensitivity and image fidelity, in support of an eventual final data release.

\subsection{Confusion limits}
\label{sec:confusion}
A fundamental limit to the sensitivity of radio (and other) images is the ``confusion'' level, the rms fluctuations in an image due to the contribution of the entire source population that is below the level where individual sources can be detected \citep{1974ApJ...188..279C}. Using the formalism detailed in \citet{2012ApJ...758...23C} and further developed by\citet{2014MNRAS.440.2791V} and \citet{2015MNRAS.447.2243V} we expect the confusion level with EMU's default $15''$ synthesised beam and 943\,MHz frequency to be $\approx 15\,\mu$Jy\,beam$^{-1}$. This is comparable to the thermal noise, as measured in the Stokes V images. Further reductions in noise would therefore result in only marginal improvements in sensitivity, so EMU is very close to being as sensitive as possible at its resolution and frequency.

The EMU total intensity images have a typical rms of $\approx\,30\,\mu$Jy\,beam$^{-1}$, a factor of $\approx\,2$ brighter than either the thermal rms or confusion level separately, or $\approx\,1.5$ more than the $\approx\,20\,\mu$Jy\,beam$^{-1}$ expected from their combination (a similar excess is found for the 151\,MHz, $20''$ data products of LoTSS). To assess the contributions to the EMU image rms from various factors, we made difference images between repeated observations of the same fields. The rms fluctuations in the difference images did not increase above that of the input images, as would be expected if it arose purely from thermal noise, which is independent for each observation. Thus, the measured rms fluctuations contain contributions both from confusion and instrumental effects such as the cumulative sidelobes from very faint sources, as well as the thermal noise. Further processing developments may allow us to push closer to the pure confusion limit.

To provide a more concrete illustration of EMU's sensitivity to low surface brightnesses, we convert it to a brightness temperature detection limit, which is given by 
\begin{equation}
    T = \frac{\lambda^2 S_{\rm lim}}{2k\Omega_b},
\end{equation}
where $\lambda = 0.318\,$m at EMU's 943\,MHz, $S_{\rm lim}$ is the limiting flux density of detected sources, $k=1.38\times10^{-23}\,$J\,K$^{-1}$ is the Boltzmann constant, and $\Omega_b$ is the restoring beam solid angle. Adopting a flux limit of $S_{\rm lim}=150\,\mu$Jy and EMU's $15''$ resolution, giving $\Omega_b=5.99\times10^{-9}\,$sr, we have EMU's limiting brightness temperature $T=0.92\,$K. The brightness temperature of nearby face on spiral galaxies is about $T\approx 1\,$K at 1.4\,GHz \citep{1998AJ....115.1693C}, and so about $T\approx 3\,$K at EMU's 943\,MHz observing frequency. The brightness temperature falls with redshift (for synchrotron sources with $\alpha=-0.7$) as $T\propto (1+z)^{-3+\alpha}$, implying that EMU will be able to detect such typical face on spiral or star forming galaxies to $z\approx 0.3-0.4$.

Given EMU's excellent $uv$ coverage, it is also sensitive over a wide range of angular scales. This enables the survey to probe much lower than the nominal confusion level at scales larger than $15''$, by first subtracting sources at the $15''$ resolution, and then convolving to larger scales. An automated process for doing this is described in \S\,\ref{sec:diffuse}, and results in rms fluctuations reaching $< 9\,\mu$Jy\,($15''$ beam)$^{-1}$, equivalent to $54\,\mu$Jy\,($45''$ beam)$^{-1}$ for structures on that scale and above.
Thus, EMU will provide extremely sensitive detections of faint extended sources, such as cluster halos and dying radio lobes and supernova remnants, that would not be otherwise possible.

\section{EMU Source Statistics}
\label{sec:cataloguestats}
Providing a complete and homogeneous source catalogue for EMU will require combination of independently observed tiles to exploit the full depth of the survey. This will account for the overlap regions between tiles as well as the northern tiles that are only observed in $5\,\text{h}$ scheduling blocks, as well as de-duplication of source detections where fields overlap (see \S\,\ref{sec:emucat}).
For a wide-area multi-year survey such as EMU, this process is most efficiently achieved once large contiguous regions of the survey footprint have been observed, and are implemented as the core stage of EMUCAT (\S\,\ref{sec:emucat}).
More immediately, when an EMU tile is observed the source finder \textit{Selavy} \citep{Whiting2012} is automatically run on the image as part of the ASKAP data pipeline. 
The resultant source catalogues for individual EMU tiles provide (a)~broad accessibility to EMU data for a range of scientific objectives prior to the availability of EMUCAT data products, and (b)~readily interpretable measurements with which to characterise the performance of the survey.

\begin{figure}[t]
    \centering
    \includegraphics[ width=\columnwidth, trim=50 50 50 0,clip]{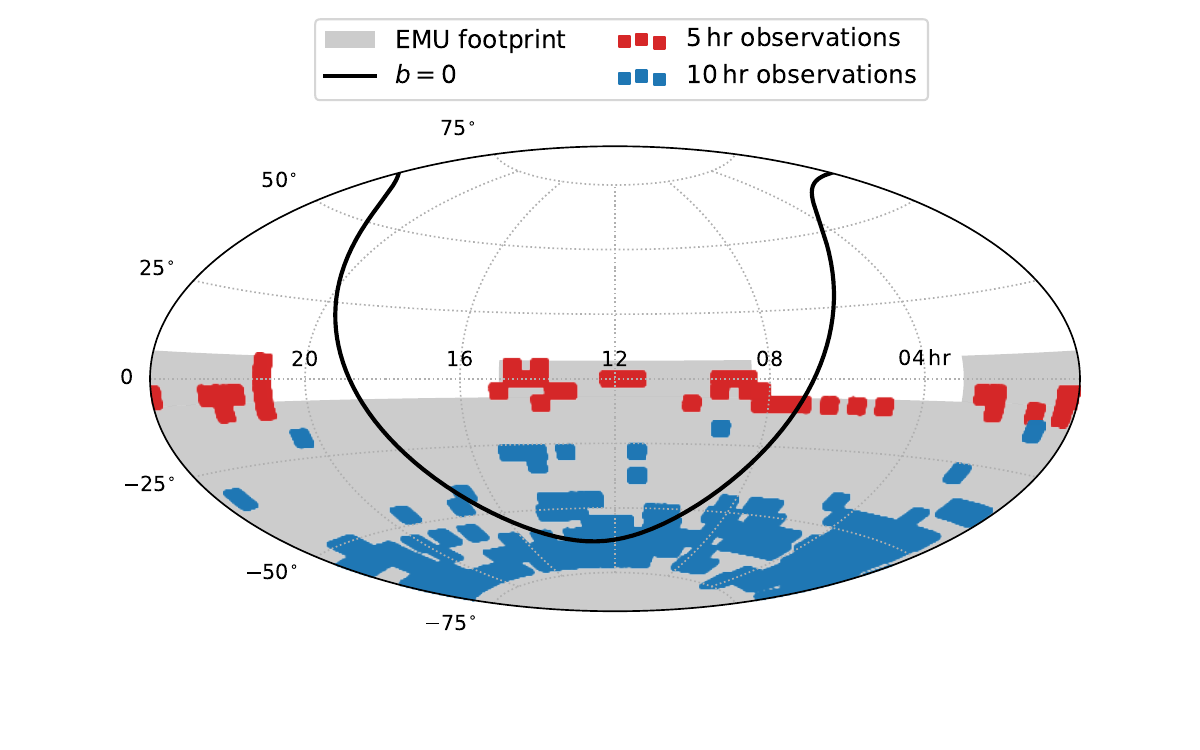}
    \caption{Sky map (Aitoff projection) showing the EMU tiles released as of July 2024.
    The grey shaded region shows the full EMU footprint, the blue regions show released $10\,\text{h}$ observations, and the red regions show released $5\,\text{h}$ observations.}
    \label{fig:emuobsaitoff}
\end{figure}

Using the second of these benefits, we can analyse the source statistics to quantify the performance of EMU to date. 
In this analysis, we draw on the data available as of July 2024. At that time 220 EMU fields ($\sim 20\,$\% of the full EMU survey) had been observed, passed validation, and were released on CASDA (Figure\,\ref{fig:emuobsaitoff}).
Of these, 143 are full-depth $10\,\text{h}$ observations and 77 are $5\,\text{h}$ observations in the northern part of the EMU footprint, 36 of which are re-observations of the same field.
We note here that the $5\,\text{h}$ observing blocks (each of which are observed twice) are combined in the image plane as part of the EMUCAT pipeline (see \S\,\ref{sec:emucat}), providing equivalent depth to the $10\,\text{h}$ observations, in order to deliver a uniform sensitivity sky catalogue.
Reimaging the fields by combining the $uv$ data for large areas of a hemispheric survey is a substantial undertaking, and is currently deferred to a later phase of the project. 
Throughout the remainder of this section we provide statistics based on the currently available automatically generated catalogues, including separate catalogues for each of the repeated $5\,\text{h}$ tiles.

The $10\,\text{h}$ observations have a median rms noise of $30\,\mu\text{Jy}\,\text{beam}^{-1}$, typically detecting $\approx 750$ sources\,deg$^{-2}$.
A few of the $10\,\text{h}$ observations are subject to increased noise, usually as a result of a bright radio source in the field, e.g., the image from SB\,46986 has an rms of $258\,\mu\text{Jy}\,\text{beam}^{-1}$ (the highest noise in any EMU tile observed prior to July 2024) a consequence of containing PKS 0131-36 \citep[a $10\,$Jy source at $843\,$MHz,][]{1992ApJS...80..137J}, but more than $95\,$\% of these observing blocks have an rms noise $< 45\,\mu\text{Jy}\,\text{beam}^{-1}$.
The $5\,\text{h}$ observations have a median rms of $46\,\mu\text{Jy}\,\text{beam}^{-1}$ (ranging between $40\,\mu\text{Jy}\,\text{beam}^{-1}$ and $60\,\mu\text{Jy}\,\text{beam}^{-1}$), detecting $\approx 460$ sources\,deg$^{-2}$ on average.
Given the overlap between adjacent EMU tiles that have been observed, simply concatenating the catalogues for individual tiles will result in some duplicate sources.
For the statistics presented here, a first-order removal of duplicates is achieved by finding sources where the distance to its nearest neighbour is less than $10''$ and where that nearest neighbour is not in the same EMU field.
For the $5\,\text{h}$ observations, the catalogue from the second observation of the tile will be dominated by the same sources that were detected in the first observation.
To avoid duplicates from these re-observations, we only include sources detected in the first of the two scheduling blocks dedicated to these tiles.

\begin{figure}
    \centering
    \includegraphics[width=0.95\columnwidth]{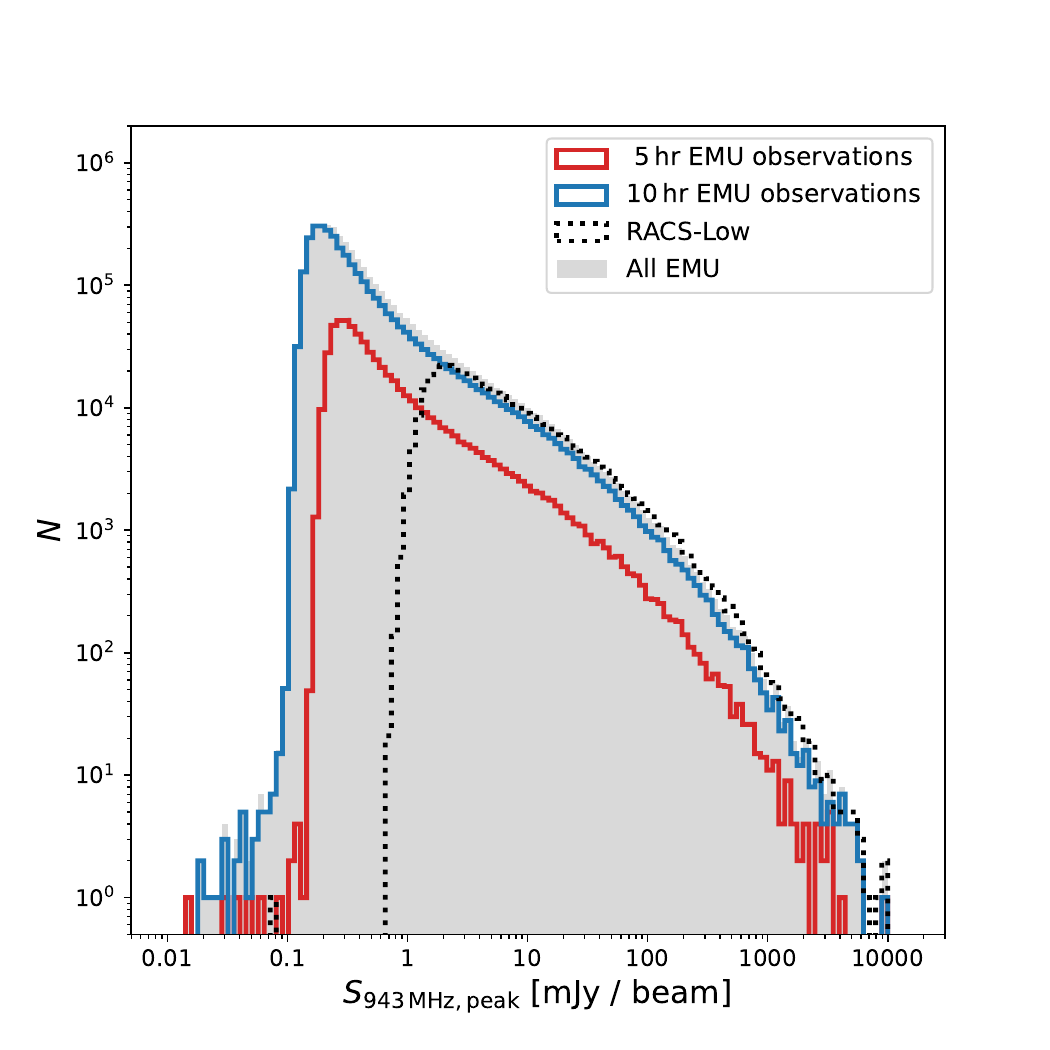}
    \caption{Distribution of the peak flux density of EMU sources. The solid grey histogram shows the EMU sources analysed here. The blue line shows sources from $10\,\text{h}$ EMU observations and the red line shows those sources from $5\,\text{h}$ EMU observations.
    For reference, the black dotted line shows the RACS low peak flux distribution normalised to account for differences in survey footprint.}
    \label{fig:fluxdist}
\end{figure}

In Figure\,\ref{fig:fluxdist} we show the peak flux distribution of all sources detected in EMU so far. 
Requiring a minimum S/N of 5, the $10\,\text{h}$ observations (blue line) detect sources down to $\approx 0.17\,\text{mJy}\,\text{beam}^{-1}$ and the $5\,\text{h}$ observations (red line) detect sources down to $\approx 0.31\,\text{mJy}\,\text{beam}^{-1}$.
For reference, we also show the peak flux distribution of sources identified by RACS in their low-frequency observations \citep[RACS-Low,][]{2020PASA...37...48M}.
RACS-Low covers $\approx 30\,000\,\text{deg}^{2}$ at $\nu\sim 888\,$MHz down to an rms noise level of $\sim300\,\mu\text{Jy}\,\text{beam}^{-1}$, and is in effect a shallower analogue of EMU.
The black dotted line in Figure\,\ref{fig:fluxdist} is obtained by randomly sampling the RACS-Low catalogue \citep{Hale2021} to normalise for the smaller observed footprint of EMU, and scaled assuming a typical spectral index of $\alpha = -0.7$ to account for the small difference in the observing frequencies of the two surveys.
Qualitatively, the bright end of the EMU flux distribution behaves as expected for a deeper version of RACS-Low with the two surveys having near identical distributions above $S_{943\,{\rm MHz}} \approx 2\,\text{mJy}\,\text{beam}^{-1}$.
The notable increase in radio source number counts at $S_{943}\lesssim 1\,$mJy can be attributed to the increasing proportion of star-forming galaxies at these faint levels \citep[e.g.,][]{1985ApJ...289..494W,1998MNRAS.296..839H,2008MNRAS.386.1695S}, highlighting the ability of EMU to probe radio source populations that are missed by shallower surveys such as RACS.

\subsection{Fraction of Resolved Sources}
\label{sec:resolved}
\begin{figure}
    \centering
    \includegraphics[width=0.95\columnwidth]{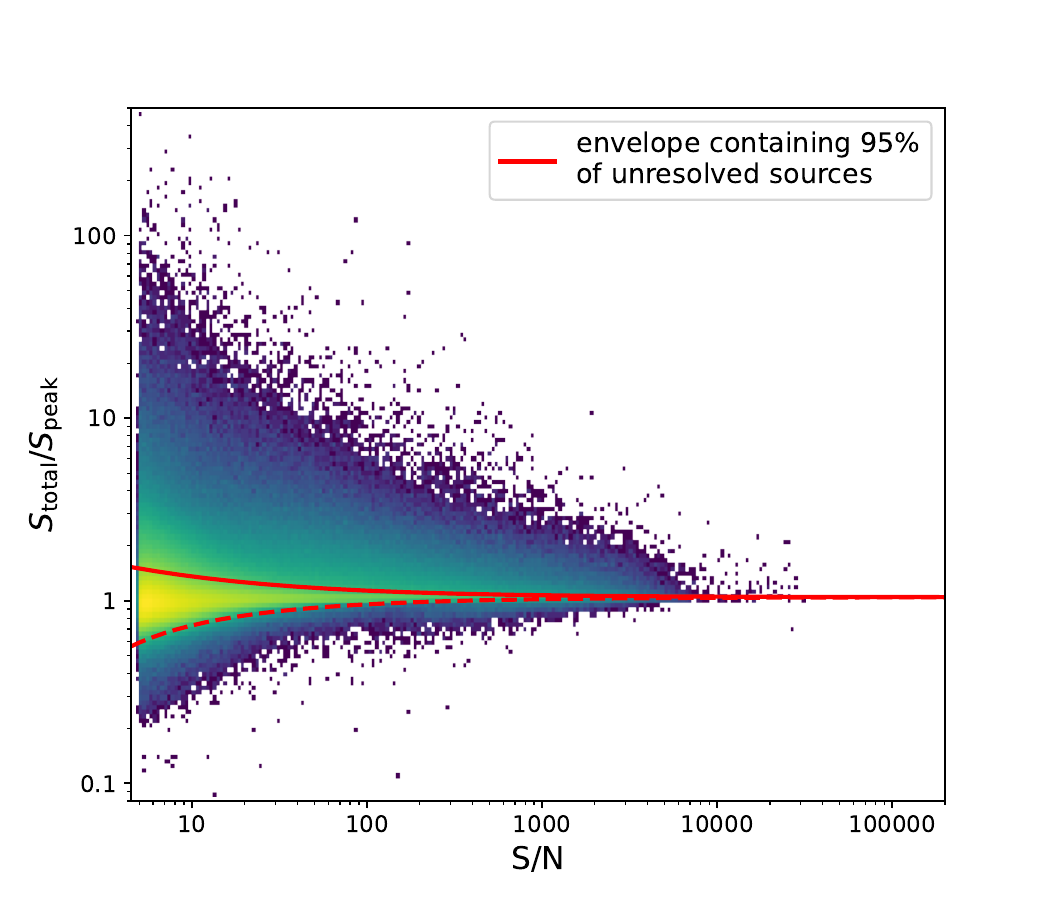}
    \caption{Distribution of total to peak flux density of EMU sources as a function of S/N. The red solid line denotes the cut off for identifying resolved sources and is a mirror of the red dashed line based on the 95\% scatter in compact sources.}
    \label{fig:tot2peak}
\end{figure}

The majority of radio sources detected by EMU are expected to be unresolved by the EMU beam \citep{2006AJ....132.2409N, 2021PASA...38...46N}.
In such cases, if the EMU point spread function is well modelled by a Gaussian, then the total flux of a point source in EMU should be equal to the amplitude of the Gaussian fitted to that source by \textit{Selavy}, at least within measurement uncertainties.
For resolved sources however, one would expect the total flux density, $S_{\text{total}}$, to be greater than the peak flux density, $S_{\text{peak}}$.
In order to identify unresolved sources in EMU, we adopt a commonly-used approach of defining an envelope in the parameter space of $S_{\text{total}}/S_{\text{peak}}$ versus S/N (Figure\,\ref{fig:tot2peak}), where the S/N is given by $S_{\text{peak}}/\text{rms}$ \citep[e.g.][]{2017A&A...602A...1S, Shimwell2019, Hale2021}.
Such an envelope is expected to be of the form:
\begin{equation}
\label{eq:totpeakenv}
\frac{S_{\text{total}}}{S_{\text{peak}}} = \text{med}\bigg(\frac{S_{\text{total}}}{S_{\text{peak}}} \bigg) \pm A\ (S/N) ^{-B}.
\end{equation}
To define our envelope, we first select a sample of likely compact and isolated sources.
In practice, we obtain these by requiring that the source is the only Gaussian fitted to that source island, and that the estimated major axis of the source is less than $20''$.
For the most robustly measured of such ``quasi compact'' sources ($S/N > 100$) we determine the median $S_{\text{total}}/S_{\text{peak}}$ to be 1.05.
To quantify the scatter in $S_{\text{total}}/S_{\text{peak}}$ as a function of $S/N$, we measure the 2.5th percentile of $S_{\text{total}}/S_{\text{peak}}$ in 10 bins of $S/N$ with equal logarithmic width.
Our lower envelope is then determined using a non-linear least-squares fit assuming the form of Equation\,\ref{eq:totpeakenv}, finding $A=1.09$ and $B=0.54$ (dashed red line in Figure\,\ref{fig:tot2peak}).
Taking the mirror of this line around the median of $S_{\text{total}}/S_{\text{peak}}$ as the upper envelope (solid red line in Figure\,\ref{fig:tot2peak}), we expect 95\,\% of the compact sources to lie in the region bounded by:
\begin{equation}
\label{eq:emutotpeakenv}
\frac{S_{\text{total}}}{S_{\text{peak}}} < 1.05 + 1.09\ (S/N) ^{-0.54}.
\end{equation}
We then find that $17\,\%$ of all the EMU sources (not just those defined as compact) have $S_{\text{total}}/S_{\text{peak}} > 1.05 \pm 1.09\ (S/N) ^{-0.54}$ and can be considered resolved, with the remaining $83\,\%$ likely being point sources at EMU's resolution.

\subsection{Astrometry}
\label{ssec:astrometry}

\begin{figure}
    \centering
    \begin{subfigure}[b]{1\linewidth}
    \includegraphics[width=0.95\columnwidth]{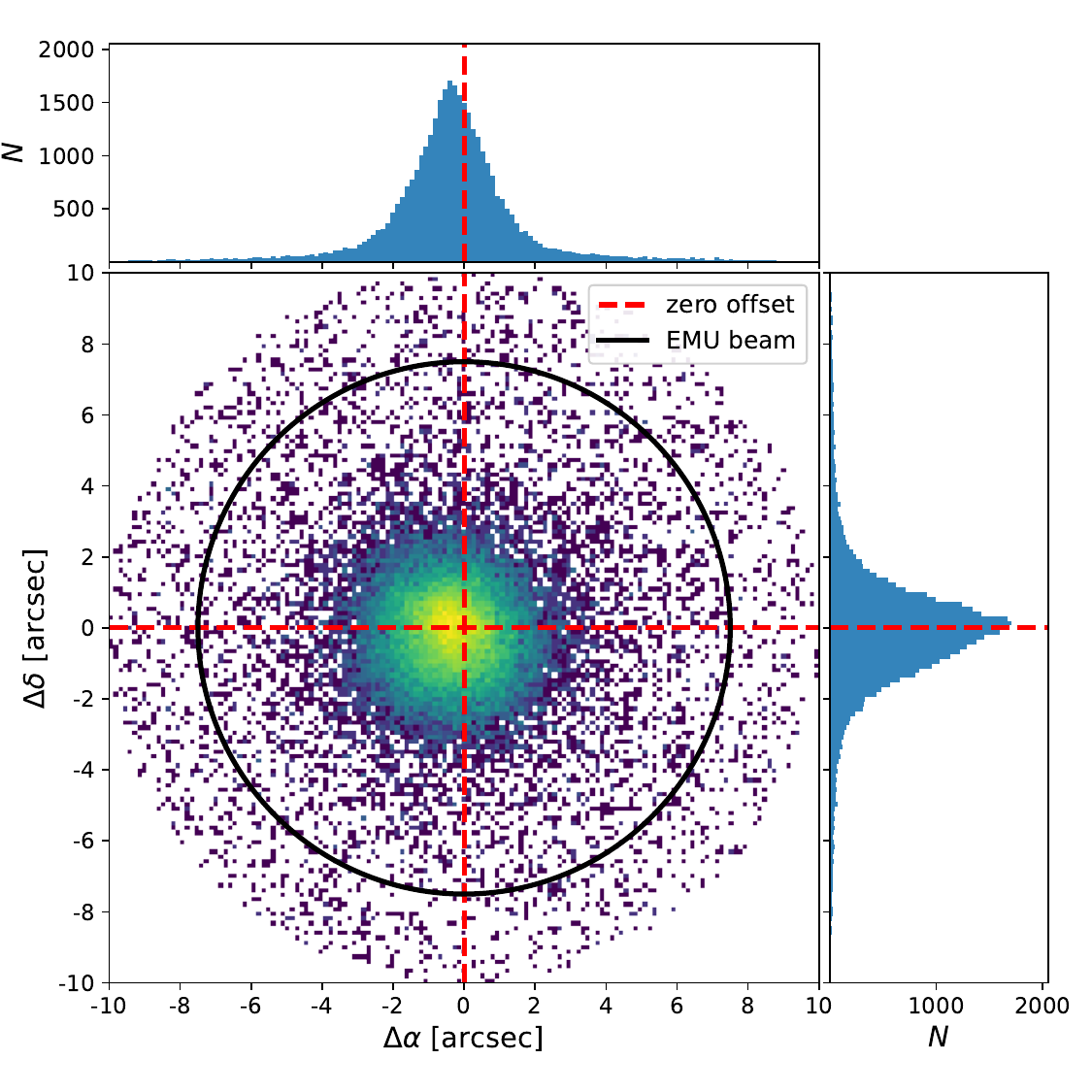}
    \caption{\label{fig:astrometry:gaia}}
    \end{subfigure}\\%
    \begin{subfigure}[b]{1\linewidth}
    \includegraphics[width=0.95\columnwidth]{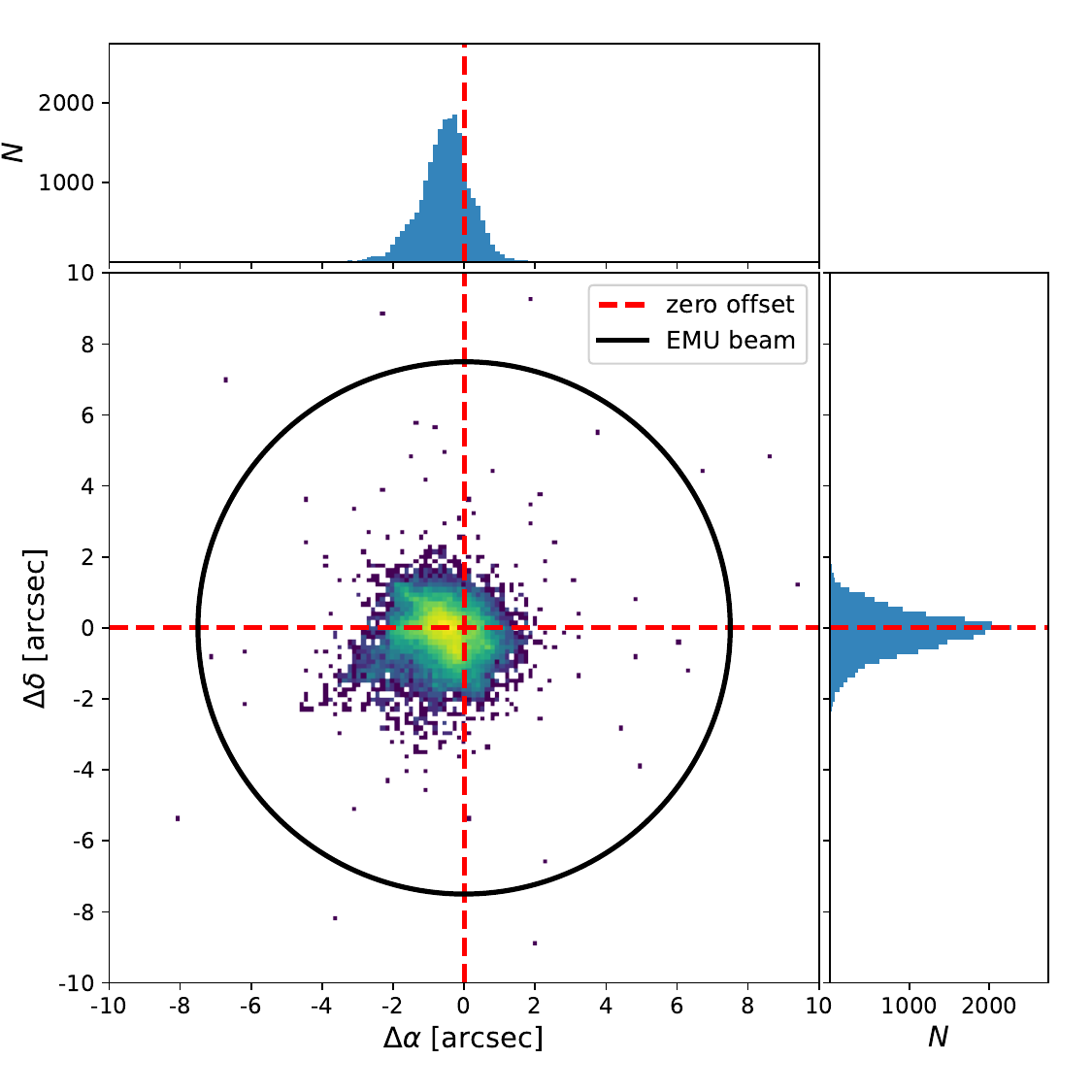}
    \caption{\label{fig:astrometry:vlass}}
    \end{subfigure}\\%
    \caption{\label{fig:astrometry} Positional offsets between EMU sources and \textit{Gaia} QSOs (top) and VLASS point sources (bottom).
    The red dashed lines show zero offset, and the black circle shows the FWHM of the typical EMU beam ($15''$).}
\end{figure}

In order to quantify the typical astrometric precision of EMU, we cross-match the point sources, i.e., those lying below the solid red line in Figure\,\ref{fig:tot2peak}, to the closest source in two other surveys, using a $10''$ search radius.
First we match EMU sources with the \textit{Gaia} third data release, a catalogue with sub-milliarcsecond astrometric precision.
\textit{Gaia} is an optical telescope, and so the vast majority of sources observed by \textit{Gaia} will not have radio counterparts.
To minimise the contamination of our cross-match by such objects, we only cross-match with \textit{Gaia} sources considered likely to be quasars ($p_{\text{QSO}} > 0.8$), finding $\approx 35\,000$ matches.
In Figure\,\ref{fig:astrometry}a we show the distribution of offsets in right ascension, $\alpha$, and declination, $\delta$, for these matches where $\Delta\alpha = (\alpha_{\text{EMU}} - \alpha_{\text{Gaia}})\cos(\delta_{\text{EMU}})$ and $\Delta\delta = \delta_{\text{EMU}} - \delta_{\text{Gaia}}$.
We find the means of the positional offsets to be $\Delta\alpha = -0.319\pm0.011''$ and $\Delta\delta = -0.109\pm0.011''$ (uncertainties are standard errors of the means), with a standard deviation in both $\Delta\alpha$ and $\Delta\delta$ of $2.14''$.

Second, we match EMU sources to the first epoch ``Quick Look'' catalogue from VLASS \citep{Lacy2020, Gordon2021}.
With a typical angular resolution of $\approx 3''$ and sub-arcsecond astrometric precision, VLASS can provide some of the most robust radio positions of any wide area sky survey.
VLASS is substantially less deep than EMU, being sensitive to point sources brighter than $\sim 1\,\text{mJy}$, and consequently one would expect the majority of VLASS sources in the survey overlap regions to be detected by EMU, significantly reducing the contamination by spurious matches that may impact our cross-match with \textit{Gaia}.
To account for the differences in resolution between the two surveys we only cross-match with likely compact sources in VLASS (deconvolved major axis of less than $1''$), finding $\approx 21\,000$ matches and showing the distribution of their positional offsets in Figure\,\ref{fig:astrometry}b.
The typical positional offsets between EMU and VLASS sources are $\Delta\alpha = -0.538\pm0.005''$ and $\Delta\delta = -0.137\pm0.005''$, with standard deviations in $\Delta\alpha$ and $\Delta\delta$ of $0.760''$ and $0.678''$, respectively.

Both our cross-match with \textit{Gaia} and our cross-match with VLASS are consistent with EMU sources being offset from the reference frame by $\sim 0.5''$ to the west and $\sim 0.1''$ to the north.
For the first EMU Pilot Survey \citet{2021PASA...38...46N} found a similar astrometric offset in declination, but report $\Delta \alpha = +0.11''$ when comparing to CatWISE.
Notably, while our early EMU main survey data cover the entire southern sky, the EMU Pilot Survey spanned the more limited declination range of $-63^{\circ} < \delta < -48^{\circ}$.
When analysing the RA bias in our cross-match with \textit{Gaia} split by the median declination of our sample ($\delta = -54^{\circ}$), those EMU/\textit{Gaia} matches at $\delta < -54^{\circ}$ have a smaller RA bias ($\Delta \alpha = -0.18\pm0.02''$) than those at $\delta > -54^{\circ}$ ($\Delta \alpha = -0.44\pm0.02''$), suggesting a potential declination dependence in the EMU astrometric bias.
The origin of this astrometric bias is currently unknown and being investigated.
Furthermore, we note that the scatter in positions we find is greater than that found in the EMU Pilot Survey \citep[see Figure 14 of][]{2021PASA...38...46N} although it is unclear whether this is due to increasing solar activity, the greater use of daytime observations, or some other cause.
The sub-arcsecond scale of the effect is substantially smaller than the typical resolution of the survey ($15''$) and the pixel size of the EMU images ($2''$), and is unlikely to be problematic for most science. 
However, until the cause is well understood, care should be taken in accounting for this known astrometric bias when matching EMU data to very high resolution multiwavelength observations (e.g. those using adaptive optics or space-based telescopes).

\subsection{Accuracy of Flux Density Measurements}

To quantify the accuracy of the flux density measurements made by EMU, we compare these measurements for point sources in EMU that are also detected in RACS-Low.
We pick RACS-Low as a comparison survey for a number of reasons.
First, like EMU, RACS-Low was observed using ASKAP and consequently any systematic biases resulting from the telescope design are expected to be similar for both surveys.
Second, the entire EMU footprint is covered by RACS-Low.
Third, RACS-Low is observed at median frequency $\nu\sim888\,$MHz, using a $288\,$MHz bandwidth, resulting in an overlap in the frequency coverage of the two surveys.
Consequently, spectral curvature should have a limited impact on the flux density measurements of non-variable sources observed by both EMU and RACS-Low.
Fourth, the accuracy of measurements from RACS-Low has already been characterised by comparison with other radio surveys, allowing the possibility of this single direct comparison with EMU.
Indeed, \citet{Hale2021} found a high level of accuracy in the RACS-Low flux density measurements, showing for sources observed in both RACS-Low and the Sydney University Molonglo Sky Survey \citep[SUMSS,][]{1999AJ....117.1578B} the ratio of flux measurements from these two surveys to be $S_{\text{RACS-Low}}/S_{\text{SUMSS}} = 1.00\pm0.16$, when accounting for spectral index.

While the native resolution of RACS-Low is similar to that of EMU ($\sim 15''$), the RACS-Low catalogue was produced from images convolved to a common resolution of $25''$ \citep{Hale2021}.
So as to not encounter issues arising from comparing flux densities of sources observed at different resolutions, we require a sample of sources that are unresolved in both EMU and RACS-Low, and isolated in EMU (the better resolution data).
For EMU we select sources satisfying Equation\,\ref{eq:emutotpeakenv} and being the only source fit to their flux island by \textit{Selavy}.
Unresolved sources in RACS-Low are identified by $S_{\text{total}}/S_{\text{peak}} < 1.025 + 0.69\ (S/N) ^{-0.62}$ \citep{Hale2021}.
We match our isolated and unresolved EMU sources to unresolved RACS-Low sources within $3''$, finding $\approx 22\,000$ sources.
This search radius is large enough that it accounts for the scatter in the EMU astrometric precision (see \S\,\ref{ssec:astrometry}).
A larger search radius however would not find many more matches, with $>90\,\%$ of those RACS-Low sources that are within $10''$ of an EMU source being separated by $<3''$.

\begin{figure}
    \centering
    \begin{subfigure}[b]{1\linewidth}
    \includegraphics[width=0.95\columnwidth, trim=0 0 0 1.2cm,clip]{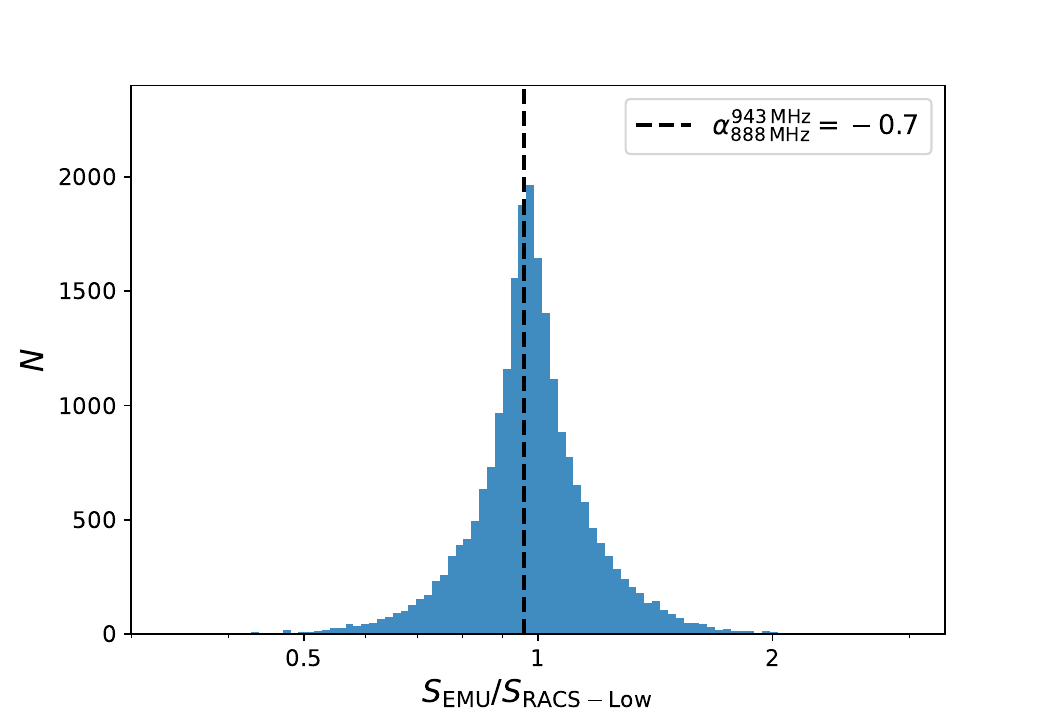}
    \caption{\label{fig:fluxratio:racs}}
    \end{subfigure}\\
    \begin{subfigure}[b]{1\linewidth}
    \includegraphics[width=0.95\columnwidth, trim=0 0 0 1cm,clip]{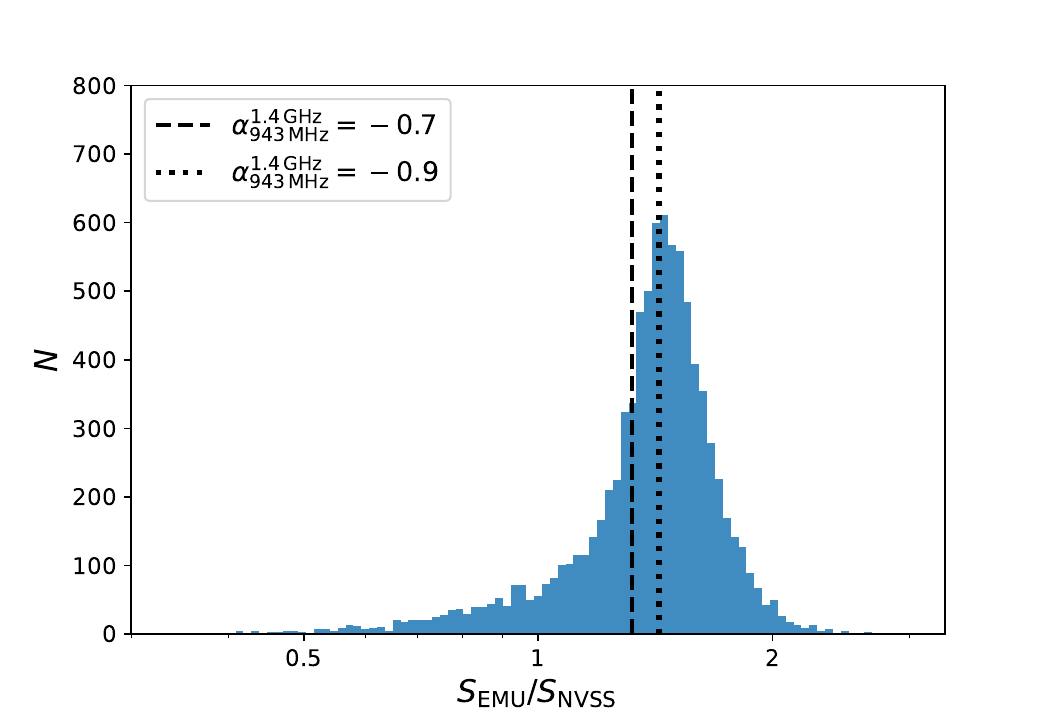}
    \caption{\label{fig:fluxratio:nvss}}
    \end{subfigure}\\
    \begin{subfigure}[b]{1\linewidth}
    \includegraphics[width=0.95\columnwidth, trim=0 0 0 1cm,clip]{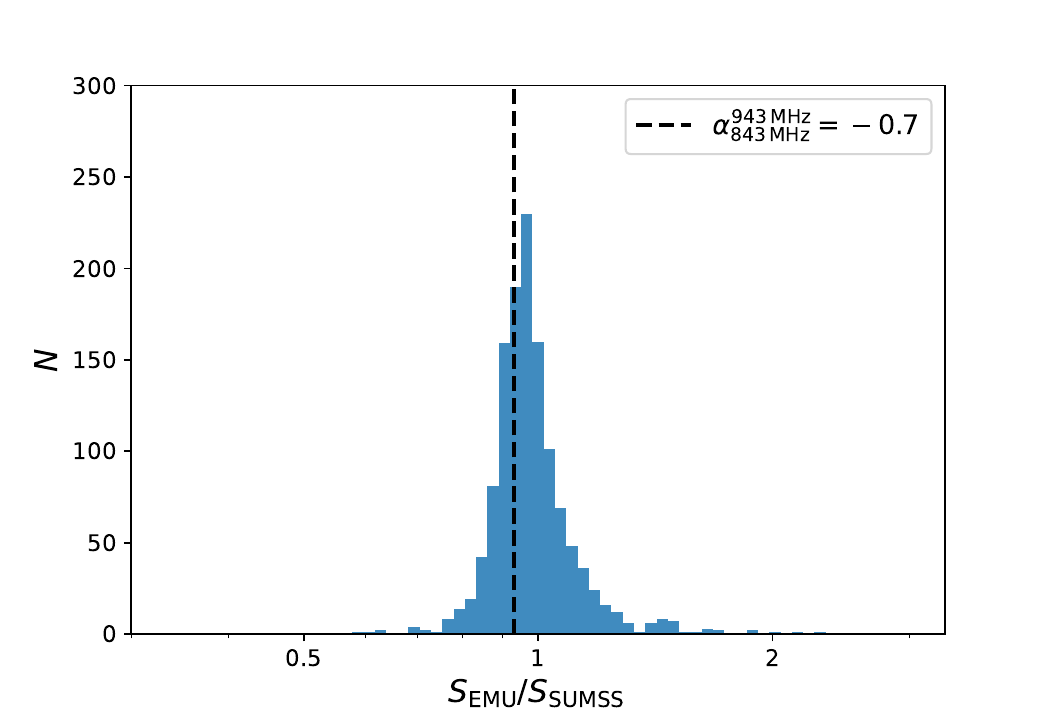}
    \caption{\label{fig:fluxratio:sumss}}
    \end{subfigure}\\
    \caption{\label{fig:fluxratio} {Distributions of the ratio of flux density in EMU to the flux density measurement in (a) RACS-Low, (b) NVSS, and (c) SUMSS, for compact and isolated sources.
    The black dashed vertical lines in all panels show the expected ratio for a source with $\alpha = -0.7$, and in panel (b) the black dotted vertical line shows the expected ratio assuming $\alpha=-0.9$.}}
\end{figure}

Figure\,\ref{fig:fluxratio} presents comparisons of the EMU flux densities against those from other surveys. In Figure\,\ref{fig:fluxratio:racs}, we show the distribution of EMU to RACS-Low flux density ratio, $S_{\text{EMU}}/S_{\text{RACS-Low}}$.
Assuming a typical two-point spectral index between the RACS-Low and EMU central observing frequencies of $\alpha_{888\,\text{MHz}}^{943\,\text{MHz}} = -0.7$ and no second-order spectral curvature, and neglecting any potential variability (the majority of EMU and RACS sources are not expected to show measurable variability), the flux density in EMU should be $96\,\%$ of the flux density in RACS-Low.
We find the median value of $S_{\text{EMU}}/S_{\text{RACS-Low}}$ to be $0.98_{-0.13}^{+0.16}$, where the uncertainties are defined by 16th and 84th percentiles.

In addition to comparing to the RACS-Low measurements, we also compare the EMU flux density measurements to sources detected by NVSS and SUMSS, two radio continuum surveys not conducted using ASKAP.
Both surveys have a poorer resolution than EMU, $45''$ for NVSS and $45''\times 45'' \text{cosec}|\delta|$ for SUMSS.
In order to ensure we are not comparing the measurements from multiple EMU sources blended into a single source in the poorer resolution surveys, we only consider EMU sources where the angular separation to their nearest neighbour in EMU is larger than the beam size of the comparison survey.
Point sources in EMU that are sufficiently isolated, by this criterion, are then matched to the catalogues of NVSS and SUMSS using a $3''$ matching radius, and the flux measurements for sources that are considered unresolved in those surveys are compared to the EMU flux density.
The NVSS catalogue does not provide both a peak and integrated flux density measurement, so we adopt those sources where the major axis of the fitted source size is smaller than or equal to or the NVSS beam size of $45''$ as being unresolved.
Similarly, SUMSS provide source sizes for resolved sources in their catalogue and we only use sources in the SUMSS catalogue they identify as unresolved \citep[see Section 4.2 of][]{2003MNRAS.342.1117M}.

In Figures \ref{fig:fluxratio:nvss} and \ref{fig:fluxratio:sumss} we show the distributions of the flux measurement ratios $S_{\text{EMU}}/S_{\text{NVSS}}$ and $S_{\text{EMU}}/S_{\text{SUMSS}}$, respectively.
In both these panels, the black dashed vertical line shows the expected ratio for a source having a typical power-law spectrum with a spectral index of $\alpha = -0.7$.
We would expect a such a source to be $32\,$\% brighter in EMU than in NVSS, which at $1.4\,$GHz is observed at a substantially higher frequency than EMU.
Instead, we find the median value of $S_{\text{EMU}}/S_{\text{NVSS}}$ to be $1.43_{-0.28}^{+0.21}$, approximately $8\,$\% larger than expected. 
There are two likely explanations for this result.
First, there could be a systematic bias in the ASKAP data that causes the flux density measurements to be too high.
Second, this result could be explained by a steeper spectral index of $\alpha = -0.9$ between $943\,$MHz and $1.4\,$GHz being more typical of the population sampled in our matches, as shown by the black dotted vertical line in Figure \ref{fig:fluxratio:nvss}, and previously suggested by \citet{Hale2021}.

The distribution of $S_{\text{EMU}}/S_{\text{SUMSS}}$ shown in Figure \ref{fig:fluxratio:sumss} can help distinguish between these two scenarios.
Like NVSS, SUMSS also used a different telescope to EMU, in this case the Molonglo Observatory Synthesis Telescope \citep{1981PASA....4..156M}, and consequently if there were a systematic bias in the ASKAP flux density measurements then we would expect the median value of $S_{\text{EMU}}/S_{\text{SUMSS}}$ to be overestimated, similarly to what we see in the $S_{\text{EMU}}/S_{\text{NVSS}}$ distribution.
The $843\,$MHz observations of SUMSS are much closer to the EMU observing frequency than the $1.4\,$GHz observations of NVSS, and consequently $S_{\text{EMU}}/S_{\text{SUMSS}}$ is less sensitive to spectral gradient (and curvature) than $S_{\text{EMU}}/S_{\text{NVSS}}$.
If the spectral index is driving the apparent overestimation of $S_{\text{EMU}}/S_{\text{NVSS}}$, then we would expect to see a smaller effect in the $S_{\text{EMU}}/S_{\text{SUMSS}}$ distribution.
Given the observing frequencies, we would expect a source with $\alpha = -0.7$ to be $93\,$\% as bright in EMU as in SUMSS.
We find the median value of $S_{\text{EMU}}/S_{\text{SUMSS}}$ to be $0.96_{-0.06}^{+0.11}$, approximately $3\,$\% larger than might be expected.
While this doesn't entirely eliminate the possibility of measurement biases in the EMU data, our results here are consistent with being driven by the spectral shape of the radio sources used, and suggest that any systematic biases in the actual ASKAP flux density measurements are typically only on the order of a few percent.

The repeated observation strategy used in the northern fields provides an opportunity to quantify the internal scatter in the EMU flux density measurements.
For any given source, the ratio of the flux density measurements taken in the second (B) observation to the first (A) observation, $S_{\text{EMU B}} / S_{\text{EMU A}}$, is a consequence of both the repeatability of the measurement, and any intrinsic source variability.
The majority of radio sources are not expected to be variable. 
Moreover, the mean time interval between the two observations of the $5\,\text{h}$ fields observed up until July 2024 is low at 18 days, and \citet{2019MNRAS.490.4898H} found that $<0.06\,$\% of radio sources vary significantly on similar timescales.
The spread in the distribution of $S_{\text{EMU B}} / S_{\text{EMU A}}$ is therefore dominated by the intrinsic measurement scatter rather than variability.
We match unresolved sources from the {\em Selavy\/} catalogues for the A and B images of the $5\,\text{h}$ fields using a $3''$ radius, finding $\approx 320\,000$ matches.
In comparing the flux measurements for these sources, we find the mean and standard deviation in $\log_{10}(S_{\text{EMU B}} / S_{\text{EMU A}})$ to be 0 and 0.06, respectively, consistent with a $\sim 15\,$\% scatter in the EMU flux density measurements.

\section{EMU VALUE-ADDED DATA}
\label{sec:data}
\subsection{EMUCAT}
\label{sec:emucat}
EMUCAT (Marvil et al., in prep) is a sophisticated resource being developed by EMU team members in coordination with the Australian SKA Regional Centre (AusSRC). The primary aims of EMUCAT are to combine all individual ASKAP EMU observations into a single cohesive all-sky data product and then add value through the inclusion of other multi-wavelength survey data and associated derived data products. EMUCAT will substantially improve the usability and scientific value of the ASKAP data and is expected to be the primary resource used by EMU science programs requiring large statistical samples. 

EMUCAT improves upon the sensitivity and uniformity of the individual EMU tile observations and provides a solution to their consolidation and de-duplication. 
An individual EMU observation, which consists of a 36-beam mosaic, has a direction-dependent sensitivity variation due to antenna primary beam and receiver sensitivity effects. EMU's survey strategy (\S~\ref{sec:survey}) includes enough overlap of adjacent tiles so as to allow for the recovery of a more uniform sensitivity when these tiles are combined. The EMUCAT workflow obtains validated tiles from CASDA and creates a larger linear mosaic of these mosaics, i.e. a ``super-mosaic.'' The workflow then runs the ASKAP source finder {\em Selavy\/} on the super-mosaic and ingests the results into a database. 

EMUCAT is constructed as a relational database and assembled via automatically triggered workflows. EMUCAT uses a set of non-overlapping region definitions to divide the full EMU survey area into 281 patches of roughly 70\,deg$^{2}$ each, and the workflow is triggered once per EMUCAT region. After ingesting the {\em Selavy} results, the workflow performs a series of post-processing and cross-identification tasks and computes various derived astrophysical properties, updating the database at each stage. The value-added data produced include identification of single and multi-component radio sources through cross-matching to the \textit{WISE} catalogues (AllWISE, CatWISE, and unWISE). About 70\% of {\em Selavy} components have counterparts within $5''$ in unWISE and CatWISE (dropping to about 50\% in AllWISE). This is supplemented through delivery of multiwavelength photometry and redshifts from a subsequent series of cross-identification stages, and deriving or measuring a range of properties (colours, sizes, morphologies, AGN/SF classification, etc.). The catalogues for cross-matching currently include VHS (DR~5), Legacy Survey (DR~10), 2MASS, 6dFGS, 2dFGRS, WiggleZ, GAMA, SDSS (DR~16), DES (DR~2), and this list is being actively expanded. Full details of the catalogue matching process, associated data structures and derived products are continuing to evolve. These will be frozen in advance of EMU public data releases, and full details will be presented in the EMUCAT description paper (Marvil et al., in prep.).

EMUCAT products are currently served through a VO-compliant TAP server hosted at the AusSRC. They are accessible through TAP-aware tools such as TOPCAT, through python notebooks at the AusSRC co-located with the data, and via a user portal for basic database queries. The super-mosaics produced by the EMUCAT workflow are made available on CASDA as EMU ``derived'' data products. Periodic EMUCAT data releases will be issued over the course of the EMU survey, for which we anticipate the first public data release around mid-2025. We expect a total of three public EMU data releases over the life of the survey, with the second release consisting of between one half and two-thirds of the final dataset, and the final release to be the full dataset.

Additionally, we have developed a comprehensive detection pipeline, \citep[RG-CAT,][]{2024PASA...41...27G}, aimed at constructing radio galaxy catalogues through the use of advanced computer vision algorithms. This pipeline follows a two-step approach. First it detects compact and extended radio galaxies along with their potential host galaxies jointly in radio (EMU) and infrared \citep[AllWISE,][]{2014yCat.2328....0C} images. This is achieved using computer vision networks built on transformers (based on convolutional neural networks), supplemented by a novel keypoint detection approach to identify host galaxies. Then it uses these predictions to build a consolidated catalogue.

The computer vision model identifies sources within an image and places bounding boxes around multiple components of extended radio sources, grouping them as individual radio galaxies. The catalogue construction pipeline (see Figure~5 of \citealt{2024PASA...41...27G}) relies on predictions from the Gal-DINO model, first introduced in RadioGalaxyNET \citep{2024PASA...41....1G}, using cutouts generated from the {\em Selavy}-based components catalogue. For each cutout, bounding boxes, categories, and infrared host predictions for the central source are used to assemble the final catalogue.

These automated approaches require a training/validation set of galaxies. For this we have used the catalogue of about 3600 double-lobed radio sources obtained from the first EMU Pilot Survey by Yew et al. (in prep.). Sources were manually identified, thus avoiding the biases introduced by automated source-finders. Each source is also manually given a morphological classification and a cross-identification to infrared and optical catalogues.

Catalogue assembly depends on the prediction confidence scores from the Gal-DINO network. For each cutout, the central source with the highest confidence score is prioritised, added to the catalogue, and removed from the {\em Selavy\/} catalogue. In cases of extended radio galaxies, multiple components are grouped as a single entry, while for compact sources, a single component is included. The final catalogue integrates both compact and extended radio galaxies. A cross-matching process is then used to link radio galaxies with their multi-wavelength counterparts from infrared (CatWISE) and optical catalogues, including DES, DESI Legacy Surveys, and SuperCosmos. Gupta et al., (in prep.) present the value added catalogues from the RG-CAT pipeline for the first year's observations of the full EMU survey.

\subsection{Diffuse Image Pipeline}
\label{sec:diffuse}
\begin{figure}[t]
    \centering\includegraphics[width=1\linewidth]{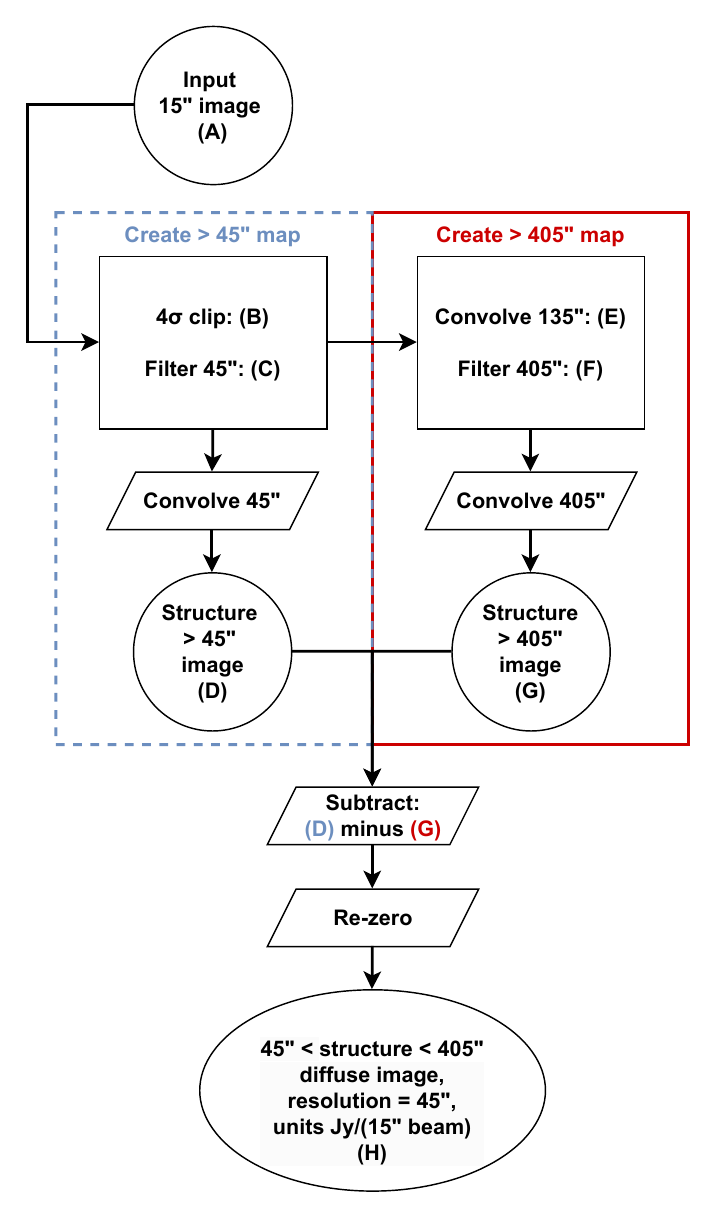}
    \caption{\label{fig:diffuse:flow} EMU diffuse imaging pipeline, which produces images with structures on scale sizes of $\sim 45''$ to $\sim 405''$. The intermediate steps are used to eliminate artifacts and improve the filtering. The script implementing this, with options for obtaining intermediate step outputs, is described in the text.}
\end{figure}

\begin{figure}[t]
    \centering
    \begin{subfigure}[b]{1\linewidth}
    \includegraphics[width=1\columnwidth]{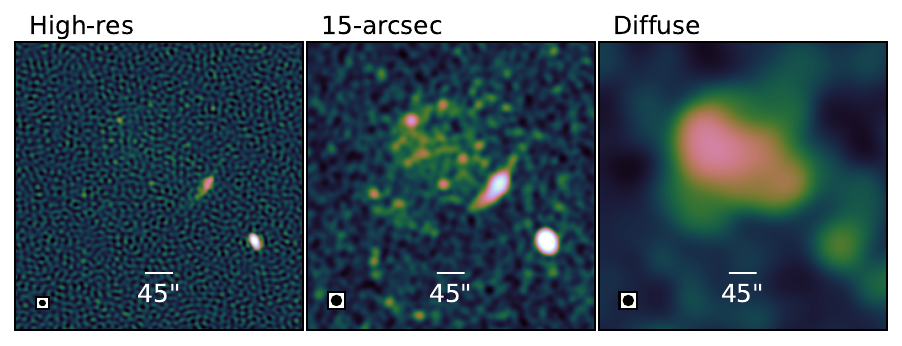}
    \caption{\label{fig:diffuse:cluster}}
    \end{subfigure}\\%
    \begin{subfigure}[b]{1\linewidth}
    \includegraphics[width=1\columnwidth]{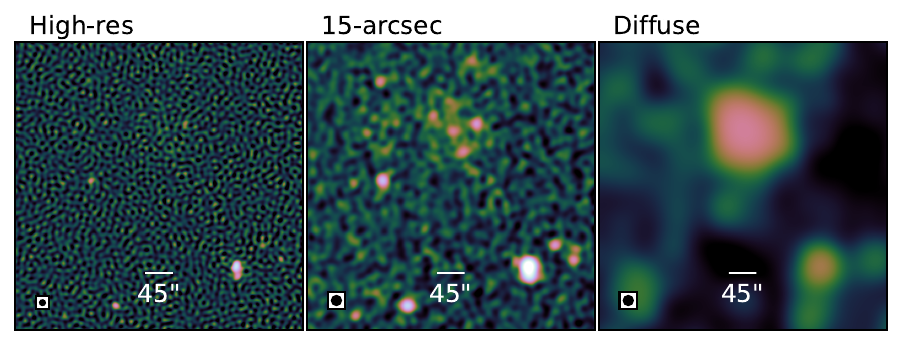}
    \caption{\label{fig:diffuse:unknown}}
    \end{subfigure}\\%
    \caption{\label{fig:diffuse:examples} Comparisons of the standard Stokes I output images for (\subref{fig:diffuse:cluster}):~the galaxy cluster ACT-CL~J0046.4$-$3911 and (\subref{fig:diffuse:unknown}):~a source of unknown origin. For each we show the high resolution image (\textit{left}), the main image at $15''$ resolution (\textit{centre}), and the diffuse image (\textit{right}). The colour scales are linear in the range $[-2\,\sigma_\text{rms}, 3\,\sigma_\text{rms}]$, and logarithmic in the range $(3\,\sigma_\text{rms}, 50\,\sigma_\text{rms}]$, where $\sigma_\text{rms}$ is the rms noise of the particular image. }
\end{figure}

The EMU data products contain not only a high density of compact and somewhat extended sources, but are also highly sensitive to more diffuse emission (see also \S\,\ref{sec:confusion}). 
This sensitivity to many angular scales can hinder detection of extremely low brightness diffuse emission, and so it becomes useful to create images containing \textit{only} the diffuse component. A common method for removing compact emission is to image data with a $(u,v)$ cut corresponding to the angular scales to be removed. The resulting CLEAN component model is then subtracted from the visibilities leaving only the extended emission. This process can be computationally expensive, requiring multiple additional imaging rounds on each visibility dataset. For the full EMU survey, we opt to employ an image-based approach to separate out diffuse emission from the embedded compact emission. This process makes use of the filtering algorithm described by \citet{Rudnick2002}. The EMU implementation of this pipeline performs two rounds of filtering that scale with the size of the restoring beam ($\theta_\text{beam}$): a lower filter removing emission $<3\,\theta_\text{beam}$ ($45''$) and an upper filter removing emission $>27\,\theta_\text{beam}$ ($405''$). A schematic outline of the pipeline is shown in Figure~\ref{fig:diffuse:flow}.

The small-scale filter removes point sources and the large-scale filter is intended to remove some of the larger features of the Galactic Plane and large-scale background ripples found in the data. Between application of the two filters, we also convolve the resulting image to $9\,\theta_\text{beam}$ ($135''$) to help reduce artifacts, although that image is not included as a final product. Finally, we reset the zero level of the images by subtracting the $\sigma$-clipped mean value over the image. We note that the zero level is not necessarily accurate at all locations on the image and the filters do not cut off sharply at the filter size. Some emission remains up to approximately three times the size of a particular filter. 
Note that this process is qualitatively very different from a simple smoothing. If the $15''$ images had merely been convolved to $45''$ resolution, the diffuse emission would have been blended with the compact emission, and their fluxes and extents would be very uncertain.

The pipeline is implemented in a \texttt{python} package \footnote{\texttt{DiffuseFilter}: \url{https://gitlab.com/Sunmish/diffusefilter}.} which also has generic implementations of the filtering algorithm described by \citet{Rudnick2002}. It provides a number of options for the user. 
Caution is advised for quantitative use of these images, and we suggest that those interested in large-scale emission in the Galactic Plane should run the filtering code without the large-scale filter applied. 

A comparison of the filtering algorithm with other filtering scales and $(u,v)$-based methods is provided by \citet{Duchesne2024}. Two example cases are shown in Figure~\ref{fig:diffuse:examples}; Figure~\ref{fig:diffuse:cluster} highlights a typical extragalactic use-case where the filter highlights diffuse emission in the galaxy cluster ACT-CL~J0046.4$-$3911 \citep{Knowles2021}, and Figure~\ref{fig:diffuse:unknown} shows an isolated unclassified region of diffuse emission at (RA, Dec) = (19:10:36.2 -71:05), with no obvious host. In the bottom example, the average S/N of the diffuse emission is $\sim 1.6$ in the $15''$ image, and rises to $\sim 5.5$ in the diffuse image.

\subsection{Source Identification and Classification}
While $\sim 80$\% of EMU sources are unresolved (\S\,\ref{sec:resolved}), the remaining $\sim\,20$\% have extended or diffuse morphologies whose study is expected to yield some of the major astrophysical insights from EMU. Traditionally, the identification and classification of sources has been carried out by manual inspection of images, but that is impractical for the large data volumes of EMU. Nevertheless, the manual identification and classification of $\sim$ 3600 galaxies with complex morphologies in the first EMU Pilot Survey \citep{2021PASA...38...46N} has been conducted by Yew et al. (in prep.), largely to provide a training set for more scalable techniques. Such techniques fall into two broad categories: ML and citizen science.

One innovative approach to classification, with the specific goal of finding complex sources with unusual morphologies, was pioneered by \citet{2019PASP..131j8007S, 2023MNRAS.521.1429S} who introduced a complexity measure to find unusual sources. An alternative approach to finding peculiar objects is to look for outliers in sources classified using autoencoders \citep{2019PASP..131j8011R} or self-organising maps \citep{2019PASP..131j8009G,2020MNRAS.497.2730G,2021A&A...645A..89M,2022PASA...39...51G}. 

\citet{Lochner2021} introduced \textsc{astronomaly}, a general purpose anomaly detection framework which uses active learning to refine the algorithm. Recently, an extension of this process, \textsc{astronomaly:\! protege} \citep{protege}, introduced a human trainer to teach the algorithm how to find a broad range of ``interesting'' sources.

\citet{2023PASA...40...44G} presented a weakly-supervised deep learning algorithm for detecting and classifying radio galaxies. Building on this, as detailed in \S\,\ref{sec:emucat} above, \citet{2023arXiv231206728G, 2024PASA...41....1G, 2024PASA...41...27G} developed object detection methods and compared convolutional neural networks and transformer based backbones to classify radio galaxies while simultaneously detecting radio bounding boxes and their infrared hosts. 
In a recent advance, we developed a multi-modal foundation model to classify EMU radio sources using text and image prompts, deployed and available to the community in EMUSE.
These developments largely focus on extragalactic sources, but ML approaches have also been successfully applied to both compact and extended Galactic sources by \citet{2016MNRAS.460.1486R, 2024PASA...41...29R}.

An alternative approach to using ML techniques is to enlist the help of thousands of citizen scientists who can classify radio morphologies and multiwavelength identifications by eye. The first such project at radio wavelengths was the EMU-driven Radio Galaxy Zoo project (\citet{2015MNRAS.453.2326B}, Wong et al., submitted) which resulted in over 100\,000 radio source classifications of radio galaxies from the FIRST and ATLAS surveys. A successor project, Radio Galaxy Zoo (LOFAR) was used at scale to generate source associations and optical identifications for LoTSS \citep{2023A&A...678A.151H}. 
This work has now been built upon by the Radio Galaxy Zoo EMU project \citep{2023MNRAS.522.2584B}, which will be highly complementary to the ML approaches. The project is now also available in Greek, Chinese, and Urdu, making it more accessible to non-English-speaking citizen scientists\footnote{\url{https://www.zooniverse.org/projects/hongming-tang/radio-galaxy-zoo-emu}}.

An important feature of all these classification schemes is that it is now recognised that classifying a galaxy as ``Type A'' or ``Type B'' is restrictive, as in general, a galaxy may have several overlapping classifications. EMU will therefore adopt the approach advocated by \citet{2021Galax...9...85R} in which sources are tagged with multiple labels, rather than being assigned to mutually exclusive classifications.

\subsection{Redshifts for EMU}
While spectroscopy remains the gold standard of redshift measurement, even the most powerful multi-object spectrograph cannot measure redshifts for more than a small fraction of the tens of millions of sources that will be catalogued by EMU. This motivates a drive to exploit photometric redshifts and other redshift estimators, and to carefully understand their accuracy and any limitations associated with spectral template availability or training sets.

The advent of large optical and infrared surveys has triggered the development of photometric redshift techniques capable of estimating the redshifts of millions of sources. 
ML plays a significant role in the estimation of these photometric redshifts. Work by \citet{2019PASP..131j8004N} demonstrates that the data used to train the ML model significantly changes the reliability of the final estimations. Where the training sample is not representative, the number of catastrophic failures the ML models produce increases significantly. Given that optically-selected samples have quite different properties (e.g., redshift distribution) from radio-selected samples, the choice of training sample is particularly important in the radio regime, and different studies have chosen different ways of dealing with the selection of appropriate training data. 

For example, \citet{2022MNRAS.512.3662D} uses as wide a range of sources as possible in their training set, but takes the additional step before training their estimation model to segment the data using a Gaussian mixture model. Individual sparse Gaussian Process regression models are then trained on each segment, allowing for the segment that most closely matches the source needing a redshift to be used as the training sample for the estimation model. As part of the effort to create a single, all-purpose redshift catalogue, \citet{2022MNRAS.512.3662D} examined their results for specific rare and under-represented populations, including sources detected in the radio continuum. They find that, when compared with the results by \citet{2019A&A...622A...3D}, the estimated photometric redshifts perform $\sim10\%$ worse for optically luminous QSOs. \citet{2022MNRAS.512.3662D} notes that the worse performance may be due in part to the use of a training set specifically made up of representative sources by \citet{2019A&A...622A...3D}, rather than a general training set segmented into best-matching sources. Creating specific training sets based on the sources being estimated is an alternative method of constructing training sets used, for example, when estimating redshifts for QSO samples \citep{2022MNRAS.512.2099C,2022MNRAS.514....1C,2023A&A...679A.101C}, and for radio continuum sources more generally \citep{2022A&C....3900557L, 2023PASA...40...39L}. Another recent ML approach combines information from the imaging together with the catalogue-level spectral energy distributions \citep{2024arXiv241107305R} to improve the performance of photo-$z$ estimation for AGNs.

Specifically, \citet{2023PASA...40...39L} has constructed training sets matching (as closely as possible) the expected properties of the EMU survey, based on data from both the northern hemisphere (with radio sources detected by the NVSS, FIRST, and LoTSS surveys with SDSS optical photometry) and the southern hemisphere (with data from ATLAS, and data taken from ASKAP observations of the Stripe\,82 equatorial field, matched with DES optical photometry). Given the differences in optical photometry used, \citet{2023PASA...40...39L} used the data from the Stripe\,82 equatorial field to find corrections between the SDSS and DES photometry. They further tested commonly-used redshift estimation algorithms, finding that simple ML algorithms can out-perform more complex algorithms, albeit without simple methods of obtaining the uncertainties and probability distribution functions (PDFs) required by the community. Future work will include incorporating better uncertainties and PDFs for the simple algorithms, investigating methods of imputing missing data, building on work started by \citet{2021mlps.confE...1L}.
The resulting work has now been applied to the first EMU Pilot Survey (Luken et al, in prep.) and will be further developed to be used routinely as part of EMUCAT.

A particular challenge arises from the lack of high-redshift radio sources that have accurately measured redshifts. This results in existing algorithms that perform poorly for high-redshift sources, even if they deliver estimates with small uncertainties at low redshift. While for some sources redshift estimates can be derived from radio continuum data alone \citep{2020MNRAS.499.3660T}, such approaches are only applicable to extended radio galaxies with broadband spectral data. It is hoped that future spectroscopic surveys, especially those specifically targeting radio continuum selected targets such as ORCHIDSS \citep{Duncan2023} and WEAVE-LOFAR \citep{Smith2016}, will improve on this, though these will have limited overlap with EMU. 

Other approaches can also help, such as the use of drop-outs to estimate the redshift of high-redshift objects \citep{2023MNRAS.519.4902S}. While this technique can only provide a redshift range, it could add significantly to the number of sources identified as high-redshift in EMU, particularly if they can be used as priors in further refining ML algorithms.

Statistical approaches, as well, can be important in constraining the population properties and evolution of the radio sources, even in the absence of individual redshift estimates for each source. The use of clustering properties to estimate redshift distributions and luminosity functions has been demonstrated for both radio \citep{2013arXiv1303.4722M} and optical \citep{2022MNRAS.509.5467K} samples. More recently, Prathap et al., (submitted) have demonstrated that radio luminosity functions can be robustly estimated from redshift distributions without the use of directly measured redshifts for individual sources.

With other facilities and programs such as WAVES, 4HS, the LSST with the Vera Rubin Observatory, the {\em Euclid\/} satellite, and others, there will be extensive redshifts, photometric and otherwise, available to complement these resources as well.

\begin{figure*}[hbt!]
    \centering
    \includegraphics[width=0.96\textwidth]{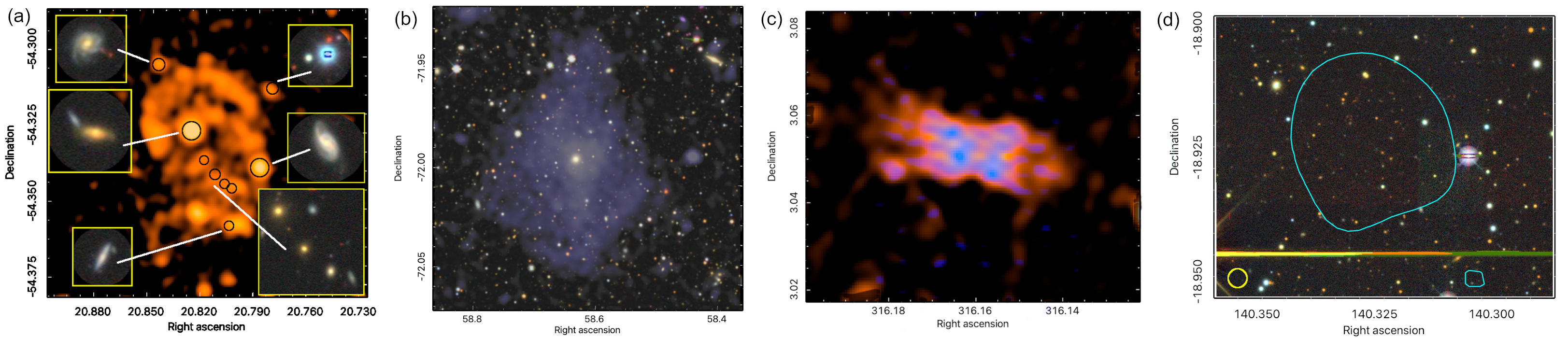}
\caption{Examples of unusual sources found in EMU. Each represents examples of physical processes that differ from current models. (a)~Potentially interacting spiral galaxies; insets drawn from the Legacy Survey. (b)~Diffuse radio emission (shown in purple) overlaid on a Legacy Survey image. (c)~EMU $\sim 15''$ radio image in grey, overlaid with $8''$ image in blue. (d)~A single contour from the diffuse image at $10\mu$Jy/15$''$ beam, outlining the location of very low brightness diffuse radio emission, overlaid on the Legacy Survey image.}
    \label{fig:selected}
\end{figure*}

\subsection{Examples of the Unexpected}
\label{sec:unexpected}
As discussed above (\S\,\ref{sec:discovery}), we expected to find sources that were rare enough, or faint enough, that they would not have been found in previous surveys. Many extended radio sources can be explained by ``weather," i.e., variations in the jets or winds or inhomogeneities in the surrounding medium. There are a small subset, though, which go beyond this. As with the ORCs discussed above, this subset contains sources that appear inconsistent with our current physical models, and have the potential to allow genuinely new insight into the associated astrophysical processes. We present four such examples here, shown in Figure~\ref{fig:selected}.

Our first example, Figure~\ref{fig:selected}a, highlights a surprising new phenomenon associated with interacting spiral galaxies. While it is well known that spiral galaxies are often radio sources, we have found larger structures surrounding some interacting spirals, resembling intersecting rings. Only one of the galaxies, J01230881$-$542015 (2MASX J01230901$-$5420130), the middle inset on the right in Figure \ref{fig:selected}a has a redshift. Adopting its value of $z=0.0729$ for the whole of the extended emission, the structure spans more than $320\,$kpc, with a luminosity of $\sim 10^{23.3}\,$W\,Hz$^{-1}$, similar to the characteristics of radio AGN.
While some of the radio emission may arise from background ellipticals in superposition, this is unlikely to be the origin of the structures seen here.
A very similar system is seen in the MGCLS survey, and at least two other examples are currently found in EMU. The origin of these extended structures associated with spiral galaxies is completely unknown, and represents new physical processes accelerating relativistic particles in their interactions.

Figure \ref{fig:selected}b shows a diffuse structure around an elliptical galaxy that is not what is expected from sources produced by pairs of jets. The host galaxy, with a $1.8\,$mJy radio core, is catalogued as WISEA J035435.09-715943.9 (2MASX J03543507-7159439), with a $J=14.38$ at $z=0.0474$. The quasi-elliptically shaped diffuse radio structure is $420 \times 315\,$kpc, with a monochromatic luminosity of $10^{23.6}\,$W\,Hz$^{-1}$, similar to radio AGN. Quasi-spherical outflows from elliptical galaxies now appear to be responsible for some ORC-like objects \citep{2024MNRAS.532.3682K}, and this diffuse structure could be a related phenomenon. Whether jetted structures could evolve into quasi-elliptical shapes such as seen here would need to be investigated through simulations.

Physically, jets from AGN fall into two broad classes. First are those where the jets maintain their collimation for long distances, dumping their energy at ``hot-spots'' and generating backflows that appear as ``lobes". Other jets expand and slowly fade in brightness as they travel away from the nucleus. The Fanaroff-Riley classification \citep{1974MNRAS.167P..31F} is one attempt to capture this dichotomy. In Figure \ref{fig:selected}c, however, we show an example of a radio galaxy that does not fit easily into either scheme. It lacks terminal hot-spots, but the extended regions near the nucleus are edge-brightened, unlike the expectations for expanding jets. There are some poorly resolved interior structures as well, with hints of a central jet. There is no obvious centrally located host, although a DESI DR10 blue disk-like galaxy with a faint W1 counterpart (WISEA J210436.49+030203.6) appears coincident with one of the radio ``patches.'' Another example of such a source, with a clear central jet, bright radio core and faint, red host galaxy, is shown in \cite{knowles22}. They argue that none of the proposed mechanisms for creating this lateral brightening appear plausible. In the case shown here, there is no obvious host galaxy. If the source is at a redshift of $z=0.5$ ($z=1$) it would have a size of $\sim 1$ ($1.3$)\,Mpc, and a luminosity of $10^{24.3}$ ($10^{25}$)\,W\,Hz$^{-1}$, making it a very large and extremely luminous radio galaxy. MHD simulations are needed to show how these sources can form, along with observed spectral structure to validate such models.

Using the diffuse images described in \S\,\ref{sec:diffuse} in combination with the $15''$ images, we can find regions of diffuse emission with no compact counterparts. Many of these show double structures with a compact radio source or bright optical galaxy between them, likely the last fading stages of a radio galaxy. There are, however, small isolated regions of diffuse emission with no likely counterparts, occuring with a surface number density approximately one per $50-100\,$deg$^{2}$. One such region is shown in Figure \ref{fig:selected}d. The contour outlines a region of diffuse emission with a brightness of $\sim 71\pm7\,\mu$Jy\,($15''$\,beam)$^{-1}$. It is also faintly visible on the original $15''$ image with a brightness of $56\pm$33$\,\mu$Jy\,($15''$\,beam)$^{-1}$ and an area of $\sim 20$ beams. The origins of these isolated diffuse patches are unclear. Among the possibilities are the one remaining lobe of a dying radio galaxy, emission from a poor cluster or group undetected in X-rays or optically, or some completely new phenomenon. Statistical studies of their optical environment and spectral indices are two areas where future studies could provide insights.

\section{EMU COLLABORATIONS}
\label{sec:discussion}

\subsection{Coordination With Related Radio Surveys}

There are three major radio survey programs that have a close relationship to the EMU survey, which provide complementary data. We describe these here.

\subsubsection{POSSUM}
\label{sec:possum}

POSSUM\footnote{POSSUM is an open collaboration that welcomes applications from qualified astronomers and students. See \text{\url{https://possum-survey.org/}}.} is a groundbreaking radio polarisation survey designed to observe commensally with EMU, sharing the same observational setup and strategy but with distinct polarimetric calibration and imaging requirements.

POSSUM aims to achieve three primary objectives: (1)~generate a dense Faraday rotation measure (RM) grid of up to one million extragalactic sources across EMU/POSSUM's survey area; (2)~map the intrinsic linearly-polarised emission and RM properties of a wide range of Galactic and extragalactic objects; and (3)~study the diffuse Galactic interstellar medium (ISM) by contributing interferometric data with excellent surface brightness sensitivity, complementing single-dish data. POSSUM will achieve an RM grid density of 30–50 RMs\,deg$^{-2}$ with a median uncertainty of $\sim$1 rad\,m$^{-2}$ and an angular resolution of $20''$ (since the observing resolution must be smoothed to a common resolution set by the lowest frequency channel of the polarimetric data cubes). POSSUM will also be supplemented by observations covering 1296–1440\,MHz over 38\% of the sky, operating commensally with the WALLABY survey, further enhancing its capabilities.

The POSSUM science case (Gaensler et al., 2025, in press) is highly complementary to that of EMU. POSSUM is focused on probing environments where magnetic fields play crucial astrophysical roles, such as AGN and radio galaxies, galaxy clusters and groups, the cosmic web and intergalactic medium (IGM), the Magellanic System and nearby galaxies, the circumgalactic medium (CGM) and galaxy halos, and the Milky Way's ISM. POSSUM will address two more fundamental questions: How were the first magnetic fields generated? And what processes have sustained, organised, and strengthened these fields through to the present day?

To leverage the complementarity between EMU and POSSUM, the primary POSSUM polarisation catalogue is designed to have a one-to-one correspondence with the EMU catalogue, ensuring that each entry in the latter contains comprehensive polarisation information from the former, whether or not the source components are significantly polarised. Additionally, POSSUM will benefit from EMU’s better image fidelity and resolution, particularly through the use of EMU’s multifrequency synthesis images smoothed to POSSUM’s resolution to create fractional polarisation maps. For resolved sources, the structural details visible in EMU’s high resolution images will aid in identifying targets and interpreting POSSUM polarisation maps, and vice versa. The EMUCAT data products from cross-matching sources with other waveband surveys and determining distance estimates will also be invaluable for interpreting POSSUM data. Furthermore, while only a small percentage of EMU sources will yield reliable individual RMs, POSSUM can study the polarisation properties of fainter sources through stacking techniques, aiding population-level studies, and making the most of the complementary datasets.

Finally, a joint project proposal and approval process has been established between EMU and POSSUM to streamline collaborative research efforts, ensuring that the scientific potential of both surveys is fully realised. The policies accommodate cases where data products from either EMU or POSSUM are used in projects led by the other survey, but where the contributions are minor enough that they do not warrant the initiation of a formal joint project.

\subsubsection{PEGASUS}
\label{sec:pegasus}

The sensitivity of ASKAP to extended emission declines at scales above a few arcminutes and drops to zero at $\approx 43'$ and $\approx 60'$ at the two ends of the EMU band, set by the shortest baseline. About $17\%$ of EMU sources have extended emission (\S\, \ref{sec:resolved}) and $\approx 0.1 - 1\%$ of extragalactic sources (based on results from the first EMU Pilot survey), as many as 20\,000--200\,000 (or about 20--200 in each tile), have extended emission on scales large enough that they could be mischaracterised. Large objects are poorly imaged and sensitivity to large spatial scales varies with frequency, impacting the reliability of spectral indices of these large objects. 

Large Galactic and extragalactic objects, such as SNRs, H\,{\sc ii} regions, planetary nebulae, Galactic ISM diffuse emission, nearby galaxy cluster intracluster medium (ICM), the Magellanic Clouds, Centaurus A, and giant radio galaxies, lose a significant fraction of their flux if measured with interferometers alone, compromising the estimate of total energy content, spectral index, and the inferred age of their relativistic electrons \citep{2012ApJ...759...25M, 2014MNRAS.445.4507I, 2017A&A...603A.122G, 2024MNRAS.528.6470M}.

To address this issue with the largest objects, the single-dish survey POSSUM EMU GMIMS All-Stokes UWL Survey (PEGASUS; Parkes project P1123; Carretti et al., in prep.) is being conducted with the CSIRO 64\,m Murriyang radio telescope at Parkes. This complements EMU by providing missing large-scale information. PEGASUS uses the Ultra Wide-bandwidth Low \citep[UWL,][]{2020PASA...37...12H} receiver to map the Stokes parameters $I$, $Q$, and $U$ of the entire southern sky up to $\delta = +20^\circ$, in the frequency range 704--1440\,MHz with a resolution of 0.5\,MHz. With a diameter of 64\,m and an angular resolution of $\approx 22'$ at the EMU centre frequency, PEGASUS data are well suited to be combined with EMU (Kothes et al., in prep.). The observations, begun in 2023, are about 70\% complete, and are expected to be finalised later in 2025.

Besides EMU, PEGASUS complements two more projects. It provides large scale information to POSSUM, described above (\S\,\ref{sec:possum}). It also completes the frequency coverage of the Global Magneto-Ionic Medium Survey (GMIMS) project in the south. GMIMS \citep[][Sun et al., submitted]{2019AJ....158...44W, 2021AJ....162...35W, 2024arXiv240606166R} is a spectropolarimetric survey to map the entire sky at 300 -- 1800\,MHz with single-dish radio telescopes and study the properties of the Galactic magnetic fields and of the magneto-ionic medium. PEGASUS is the Mid-Band (704 -- 1440\,MHz) South survey of GMIMS.
The Low-Band (300 -- 480\,MHz) \citep{2019AJ....158...44W} and High-Band (1300 -- 1800\,MHz) South surveys \citep[][Sun et al., submitted]{ 2024arXiv240606166R} have already been completed.

\subsubsection{WALLABY}

\begin{figure*} 
\centering
    \includegraphics[width=17cm]{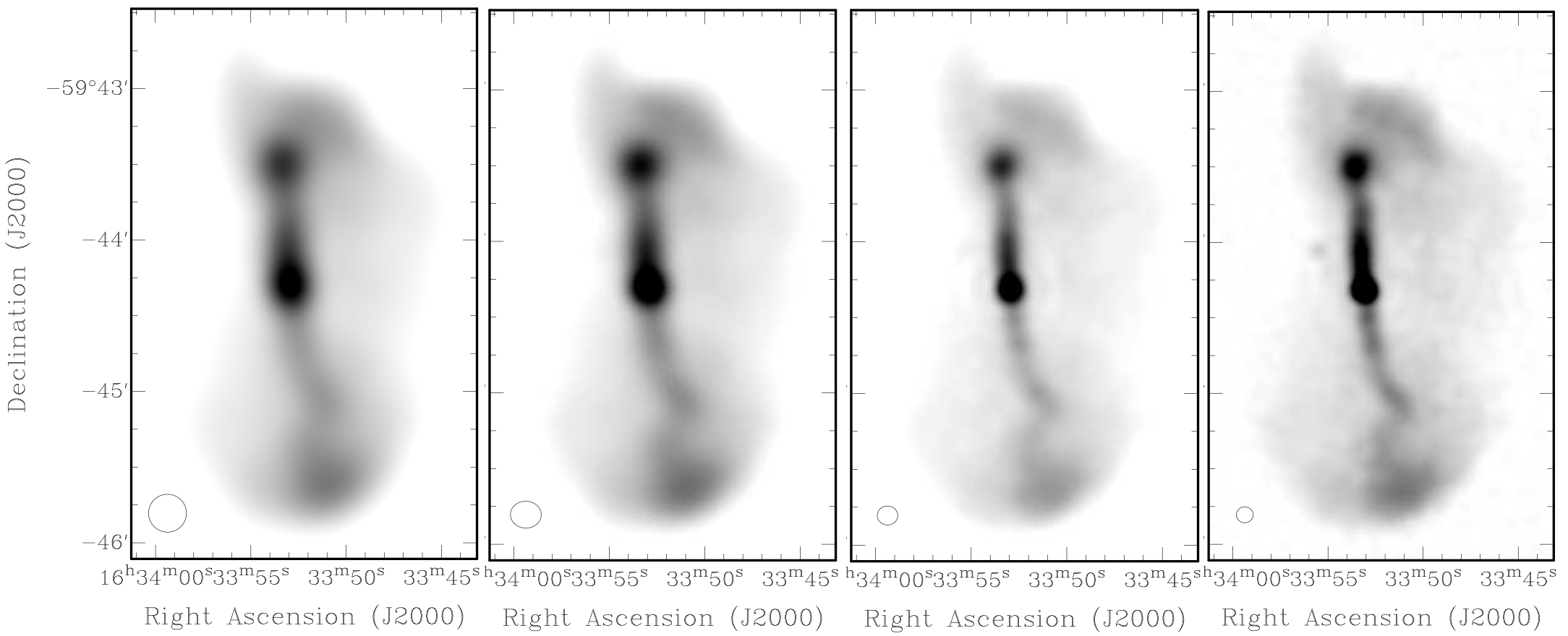}
    \includegraphics[width=17cm]{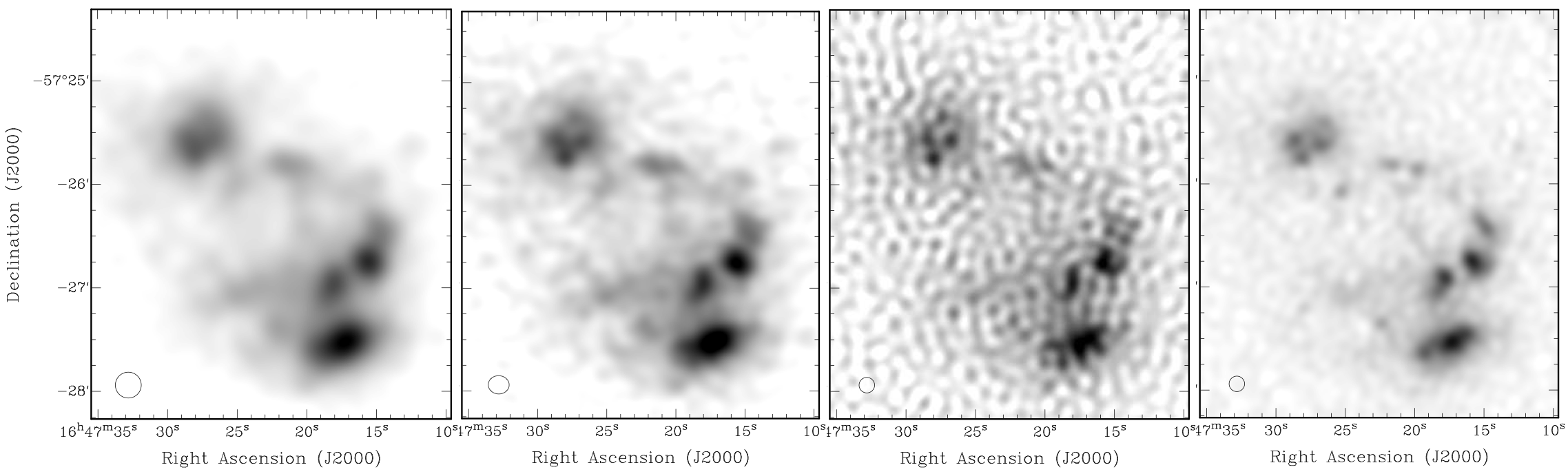}
\caption{ASKAP radio continuum images of (top) a double-lobe radio galaxy and (bottom) ESO\,179-IG013, also known as Kathryn's Wheel \citep{2015MNRAS.452.3759P,2024ApJ...967L..26P}. The first three images left to right are from EMU, showing the ``conv'', ``raw'', and ``highres'' imaging at 943\,MHz respectively. The fourth (rightmost) image is from WALLABY at 1.4\,GHz. The strengths and limitations of the EMU ``highres'' data, and the value of WALLABY continuum imaging as a complement, can be seen in particular in the Kathryn's Wheel example.
}
    \label{fig:DRG-collage}
\end{figure*}

\label{sec:wallaby}
Another ASKAP sky survey that delivers deep radio continuum data is WALLABY \citep{Koribalski2012,2020Ap&SS.365..118K}. While its main purpose is to map neutral atomic hydrogen (H\,{\sc i}) in and between galaxies, it also delivers 1.4\,GHz radio continuum images. WALLABY was allocated a total of 8832\,h over five years \citep{Westmeier2022}. With an integration time of 16\,h (typically $2 \times 8$\,h) per field, the WALLABY rms is $\sim 1.6\,$mJy\,beam$^{-1}$ per 4\,km\,s$^{-1}$ channel for the H\,{\sc i} 21\,cm spectral line ($\sim 30''$ resolution) and $\sim 30\,\mu$Jy\,beam$^{-1}$ for the radio continuum ($\sim 8''$ resolution). WALLABY was initially proposed in 2008 as a 21\,cm sky survey including H\,{\sc i} spectral line, continuum and polarisation data. Due to RFI, only half the available ASKAP bandwidth (i.e, 144\,MHz) is currently being processed, providing images over the 1300 -- 1440\,MHz frequency range. As of April 2025, only $\sim 6\%$ (62) of the 1104 WALLABY fields have been observed and validated as good, due to a delayed survey start and solar avoidance constraints.

WALLABY data products include H\,{\sc i} images, spectra and redshifts of galaxies in the nearby Universe (resolution $30''$, 4\,km\,s$^{-1}$) as well as 1.4\,GHz radio continuum images (resolution $\sim 8''$, see Figure~\ref{fig:DRG-collage}). It is worth noting in this comparison that the sensitivity cannot be compared between the EMU ``high-res'' image and the WALLABY image, although both have similar resolution. The EMU high-res image is not optimised for sensitivity, given it is constructed with uniform weighting, which optimises resolution at the expense of sensitivity. The EMU ``conv'' image has a comparable sensitivity, from a shorter integration time, but poorer resolution due to the lower observing frequency.
High resolution ($12''$) H\,{\sc i} cut-outs are also created for pre-selected gas-rich galaxies \citep{2020Ap&SS.365..118K, Murugeshan2024}, selected from the H\,{\sc i} Parkes All-Sky Survey \citep[HIPASS,][]{Koribalski2004,Meyer2004}. An example multi-wavelength image is shown in \citet[][their Figure~8]{2011PASA...28..215N} highlighting the inner, star-forming disk of the spiral galaxy M\,83.

Radio continuum images from EMU and WALLABY together deliver extended frequency coverage, useful to determine spectral indices \citep[e.g.,][]{2024MNRAS.533..608K}, and being used jointly in deriving POSSUM data products (Gaensler et al., 2025, in press).

\section{SUMMARY AND NEXT STEPS}
\label{sec:conclusion}
The Evolutionary Map of the Universe will be the touchstone radio atlas of the southern hemisphere. It will deliver imaging with a median sensitivity of $\sigma = 30\,\mu$Jy\,beam$^{-1}$ and a resolution of $15''$ with a $288\,$MHz bandwidth centred at a frequency of $943\,$MHz. Source catalogues and value-added data products will be produced for an estimated 20 million extragalactic sources, along with high fidelity imaging of the Galactic Plane and the Magellanic Clouds. This resource is already proving invaluable in supporting scientific goals spanning galaxy evolution, galaxy environment and large-scale structure including clusters, the astrophysics of supermassive black holes, star formation and stellar evolution, the ISM, IGM, and ICM, cosmology, and much more.

The EMU collaboration consists of more than 400 researchers internationally, from 24 different countries. The collaboration is open to adding new members who are willing to contribute to team activities, and to abide by team policies. The EMU publication policy\footnote{\url{http://askap.pbworks.com/w/page/140981337/EMU Publication Policy}} has been developed to encourage collaboration within and between teams, and aims to be flexible in order not to stifle projects and collaborative work. Those wishing to join should contact the EMU management team\footnote{emu\_mt@mq.edu.au} outlining their interest in joining the collaboration, what they expect to contribute, and that they have reviewed and agree to abide by the EMU publication policy.
Publicly available ASKAP data, including EMU data, are available through CASDA. EMU is identified as project AS201, and a search on this project number will return all publicly available EMU data.
EMU publications are being tracked through an ADS library\footnote{\url{https://ui.adsabs.harvard.edu/user/libraries/H_YrZLcBTmmw3vGcki-eog}}.

The legacy value of EMU is in its large and well-characterised dataset. Well-characterised survey projects such as EMU provide an enormously valuable resource for future work that extends into areas and domains beyond those originally envisaged. As the most sensitive GHz survey to span the whole of the southern hemisphere for the foreseeable future, EMU's legacy value is expected to be vast, in line with the high impact of earlier generations of large sky surveys \citep[e.g.,][]{1995ApJ...450..559B,1998AJ....115.1693C,1999AJ....117.1578B,2003AJ....126.2081A}.

EMU tiles and {\em Selavy} catalogue products are publicly available on CASDA, listed under ``Project AS201''. The value-added EMUCAT data are proprietary to the EMU team until distributed through the regularly planned public data releases. These products include the super-mosaics and diffuse images, full-sensitivity de-duplicated catalogues and other catalogue cross-matched and derived data products. EMU observations with ASKAP will continue into 2028 at which stage this set of observations is expected to be completed. We anticipate that a future phase of the project will extend the sky coverage to $\delta = +30^{\circ}$, in order to deliver on the full potential of such a radio atlas with ASKAP, and to complement other wide-area surveys including VLASS, and LoTSS, along with current and future multiwavelength resources. Such an extension will add substantial value to the cosmology science cases and others, as well as enhancing the already significant legacy value of the project.

\begin{acknowledgement}
This scientific work uses data obtained from Inyarrimanha Ilgari Bundara, the CSIRO Murchison Radio-astronomy Observatory. We acknowledge the Wajarri Yamaji People as the Traditional Owners and native title holders of the Observatory site. CSIRO’s ASKAP radio telescope is part of the Australia Telescope National Facility (https://ror.org/05qajvd42). Operation of ASKAP is funded by the Australian Government with support from the National Collaborative Research Infrastructure Strategy. ASKAP uses the resources of the Pawsey Supercomputing Research Centre. Establishment of ASKAP, Inyarrimanha Ilgari Bundara, the CSIRO Murchison Radio-astronomy Observatory and the Pawsey Supercomputing Research Centre are initiatives of the Australian Government, with support from the Government of Western Australia and the Science and Industry Endowment Fund.

This paper includes archived data obtained through the CSIRO ASKAP Science Data Archive, CASDA (\url{http://data.csiro.au}).

M.V. acknowledges financial support from the Inter-Uni\-vers\-ity Institute for Data Intensive Astronomy (IDIA), a partnership of the University of Cape Town, the University of Pretoria and the University of the Western Cape, and from the South African Department of Science and Innovation's National Research Foundation under the ISARP RADIOMAP Joint Research Scheme (DSI-NRF Grant Number 150551) and the CPRR HIPPO Project (DSI-NRF Grant Number SRUG22031677).
R.C. acknowledges support from FCT through Fellowship PD/BD/150455/2019 (PhD: SPACE Doctoral Network PD/00040/2012) and POCH/FSE (EC). D.B. acknowledges support from FCT through Fellowship UI/BD/152315/\-2021. J.A., R.C. and D.B. acknowledge support from FCT through research grants UIDB/04434/2020 and UIDP/04434/\-2020 (DOI: 10.54499/UIDB/04434/2020 and DOI: 10.54499/\-UIDP/\-04434/2020). M.B. acknowledges funding by the Deutsche Forschungsgemeinschaft (DFG) under Germany's Excellence Strategy -- EXC 2121 ``Quantum Universe" -- 390833306 and the DFG Research Group ``Relativistic Jets''. C.L.H. acknowledges support from the Oxford Hintze Centre for Astrophysical Surveys which is funded through generous support from the Hintze Family Charitable Foundation. C.S.A. acknowledges funding from the Australian Research Council in the form of FT240100498. C.J.R. acknowledges financial support from the German Science Foundation DFG, via the Collaborative Research Center SFB1491 ``Cosmic Interacting Matters – From Source to Signal''. M.D.F. and S.L. acknowledge the Australian Research Council funding through grant DP200100784.

\end{acknowledgement}


\bibliography{EMU_main}



\end{document}